\documentclass[aps,prd,preprint,superscriptaddress,amsmath,amssymb,floatfix]{revtex4-2}
\usepackage{amsmath,amsfonts,latexsym,amssymb,graphicx,graphics,epsfig,subfigure,color,makeidx,cases}
\usepackage{xcolor,diagbox,enumitem}
\usepackage{multirow}
\usepackage[colorlinks,linkcolor=blue,anchorcolor=blue,citecolor=black,urlcolor=blue]{hyperref}
\usepackage{mathrsfs}
\renewcommand{\thesubfigure}{(\roman{subfigure})}
\makeatletter
\renewcommand{\@thesubfigure}{\thesubfigure\hskip\subfiglabelskip}
\makeatother 

\begin{document}
\raggedbottom
\title{Axial Mode Stability and Quasinormal Modes of Charged Kiselev Black Holes in Rastall Gravity}

\author{Yan Wang}
\email{wangyan2025@lzu.edu.cn}
\affiliation{
Lanzhou Center for Theoretical Physics,
Key Laboratory for Quantum Theory and Applications of the Ministry of Education,
Key Laboratory of Theoretical Physics of Gansu Province,
School of Physical Science and Technology,
Lanzhou University, Lanzhou 730000, China
}
\affiliation{
Institute of Theoretical Physics \& Research Center of Gravitation,
Lanzhou University, Lanzhou 730000, China
}

\author{Yun-tao Gu}
\email{guyt2024@lzu.edu.cn}
\affiliation{
Lanzhou Center for Theoretical Physics,
Key Laboratory for Quantum Theory and Applications of the Ministry of Education,
Key Laboratory of Theoretical Physics of Gansu Province,
School of Physical Science and Technology,
Lanzhou University, Lanzhou 730000, China
}
\affiliation{
Institute of Theoretical Physics \& Research Center of Gravitation,
Lanzhou University, Lanzhou 730000, China
}

\author{Wen-Di Guo}
\email[Corresponding author: ]{guowd@lzu.edu.cn}
\affiliation{
Lanzhou Center for Theoretical Physics,
Key Laboratory for Quantum Theory and Applications of the Ministry of Education,
Key Laboratory of Theoretical Physics of Gansu Province,
School of Physical Science and Technology,
Lanzhou University, Lanzhou 730000, China
}
\affiliation{
Institute of Theoretical Physics \& Research Center of Gravitation,
Lanzhou University, Lanzhou 730000, China
}
\begin{abstract}
We investigate axial gravitoelectromagnetic perturbations of charged
black holes surrounded by dust-like Kiselev-type anisotropic matter
in Rastall gravity. We derive the axial matter compatibility condition
and construct a compatible response by holding the covariant
anisotropy-direction vector unperturbed. The resulting gravitational
and electromagnetic perturbations obey two coupled Schr\"odinger-type
master equations. An $r$-independent algebraic decoupling exists in
the Reissner--Nordstr\"om and general-relativistic dust-like limits;
additional radial structure obstructs it in generic charged Rastall
backgrounds with nonzero surrounding matter.
We prove axial mode stability under the adopted closure for every nonextremal background 
satisfying the stated exterior conditions, covering all physical axial multipoles.
A matrix $S$-deformation excludes
exponentially growing coupled modes for $\ell\geq2$, while the
physical electromagnetic dipole has a positive effective potential.
A Chebyshev pseudospectral calculation, validated against the
Reissner--Nordstr\"om spectrum and spectral convergence, determines
the fundamental frequencies. In the parameter ranges examined,
increasing the Rastall coupling lowers both the oscillation frequency
and the damping rate in the small-coupling region; in the large-coupling
region, the oscillation frequency rises while the damping rate varies
nonmonotonically. A larger charge generally raises the oscillation
frequency, whereas stronger surrounding-matter contributions produce
longer-lived modes.
\end{abstract}

\maketitle
\section{Introduction} \label{Introduction}

Black-hole perturbation theory provides a fundamental framework for
studying the response of black holes to external disturbances
\cite{Regge:1957td,Zerilli:1970se}. After
an initial transient stage, the evolution of a perturbed black hole is
generally dominated by a discrete set of damped oscillations known as
quasinormal modes (QNMs)
\cite{Vishveshwara:1970zz,Chandrasekhar:1975zza}.
Their complex frequencies are determined by the background spacetime
and the boundary conditions at the event horizon and the outer
boundary. The real part characterizes the oscillation frequency,
whereas the imaginary part determines the damping or growth rate
\cite{Konoplya:2011qq,Berti:2009kk}.
QNMs therefore play an important role in black-hole stability and in
the ringdown stage of gravitational-wave signals
\cite{Echeverria:1989hg}, and they provide the
theoretical basis for black-hole spectroscopy
\cite{Dreyer:2003bv,Berti:2005ys}.
The theoretical role of QNMs is increasingly complemented
by strong-field observations.
Gravitational-wave ringdown measurements have begun to probe
the QNM spectrum and the Kerr nature of merger remnants
\cite{Isi:2019aib,LIGOScientific:2025wao},
whereas horizon-scale images of M87* and Sgr~A* provide
an independent probe of black-hole geometries
\cite{EventHorizonTelescope:2020qrl,EventHorizonTelescope:2022xqj}.
Together, these observations motivate a careful assessment
of how black-hole charge and environmental matter affect
the spacetime and its perturbations.

The perturbation problem becomes richer when the black hole carries an
electric charge or is embedded in a nonvacuum environment. In the
Reissner--Nordstr\"om (RN) spacetime, metric and Maxwell perturbations form a
coupled gravitoelectromagnetic system, whose spectrum can be organized
into gravitational-type and electromagnetic-type branches
\cite{Zerilli:1974ai,Leaver:1990zz,Moncrief:1974gw,Moncrief:1975sb}.
Moreover, surrounding matter can modify both the background geometry
and the perturbation spectrum
\cite{Barausse:2014tra,Leung:1997was,C:2024cnk}.
In modified gravity, coupling between perturbation sectors can
also reshape the spectral response to environmental disturbances.
For example, a study in dynamical Chern--Simons gravity found
that gravito-scalar coupling modifies the QNM branch structure
and dominant-mode transitions induced by a localized perturbation
of the effective potential
\cite{Hu:2025efp}.
A widely used phenomenological description of such
environments is the Kiselev-type construction
\cite{Kiselev:2002dx,Visser:2019brz}.
The Kiselev-type source is anisotropic. Here, \textit{dust-like} denotes
a vanishing average equation-of-state parameter, with distinct radial
and tangential pressures \cite{Hadi:2024idb}.

More broadly, alternative theories of gravity provide
several routes for exploring deviations from
general relativity.
Representative examples include curvature-based
\(f(R)\) gravity and torsion-based \(f(T)\)
teleparallel gravity
\cite{Buchdahl:1970ynr,DeFelice:2010aj,Starobinsky:1980te,Cai:2015emx}.
Extra-dimensional and thick-brane scenarios provide
another setting in which the structure and perturbative
dynamics of gravitating systems have been investigated
\cite{Rubakov:1983bb,Gremm:1999pj,Jia:2024pdk,Jia:2026vnx,Li:2026dfy,Chen:2020zzs}.
Against this broader background, Rastall gravity takes
a different phenomenological route by modifying the usual
covariant conservation law and relating the divergence
of the energy-momentum tensor to the gradient of
the Ricci scalar
\cite{Rastall:1972swe,Rastall:1976uh}.
The interpretation of Rastall gravity and the extent
to which it represents a theory dynamically distinct
from general relativity remain under discussion
\cite{Visser:2017gpz,Darabi:2017coc}.
The standard Einstein equations are recovered when
the Rastall parameter vanishes.
For convenience, we introduce the dimensionless coupling
\(\gamma\equiv\kappa\lambda\).
Within this framework, charged black-hole solutions
surrounded by Kiselev-type matter have been constructed
\cite{Heydarzade:2017wxu}.
Their properties and quasinormal spectra have subsequently
been investigated in a variety of settings
\cite{Cai:2020igv,Zhang:2024bov,MoraisGraca:2017hrf,Liang:2018tpt,Gogoi:2021dkr,Wang:2025wbu,Sarkar:2025fjx,Karmakar:2023cwg,Shao:2020gwr}.
The scope of Rastall-gravity studies extends beyond
four-dimensional black holes.
Gravitational-wave polarization modes have been analyzed
in scalar-tensor-Rastall gravity
\cite{Fan:2024pex}.
Separately, Rastall thick-brane configurations have been
investigated through their quasibound and quasinormal spectra
\cite{Tan:2024url,Huang:2026cml}.
At the phenomenological level, constraints on Rastall
parameters have also been obtained from cosmological
observations, galaxy rotation curves, and galaxy-scale
strong gravitational lensing
\cite{Akarsu:2020yqa,Tang:2019dsk,Li:2019jkv}.

Existing studies of QNMs in nonvacuum black-hole spacetimes have often
considered test fields on fixed backgrounds
\cite{Chen:2005qh,Hu:2019cml,Zhang:2006hh} or adopted approximations
in which perturbations of the external matter source are neglected
\cite{Zhang:2021bdr,Zhang:2022roh,Zhao:2023tyo,Cai:2020igv,Zhang:2024bov}.
These approaches provide a basis for understanding how surrounding
matter influences black-hole QNMs. For the coupled axial perturbations
of a charged black hole, the metric, electromagnetic field, and
surrounding matter must satisfy the linearized field equations
together. In an anisotropic-matter description, the matter response
is specified through closure relations
\cite{Cardoso:2022whc,Chen:2019iuo,Konoplya:2024lch}.
The linearized Rastall nonconservation relation supplies a compatibility
condition linking the matter and metric perturbations. Establishing
this condition and incorporating a compatible matter response are
central to constructing the coupled perturbation system.

In this work, we study axial perturbations of a charged black hole
surrounded by dust-like Kiselev-type anisotropic matter in Rastall
gravity. We derive the matter compatibility condition before fixing
the anisotropy-direction perturbation. We then adopt the closure
$\delta s_\mu=0$, obtain the compatible matter response, and reduce
the gravitational and electromagnetic equations to two coupled
Schr\"odinger-type master equations.
The standard RN system is recovered when the surrounding matter
is removed. We determine the conditions for an $r$-independent
algebraic decoupling: it exists in the RN and general-relativistic
dust-like limits, while additional radial structure obstructs it
in generic charged Rastall backgrounds with nonzero surrounding
matter.

We establish axial mode stability of the coupled system under this
closure for the nonextremal backgrounds specified below.
A constant rescaling makes the potential matrix symmetric, and an
explicit matrix $S$-deformation yields a positive-definite quadratic
form in the static exterior for $\ell\geq2$. The physical
electromagnetic dipole obeys an autonomous equation with a positive
potential and induces a constrained metric response.
Together, these results exclude exponentially growing modes in all
physical axial multipoles, independently of the numerical parameter
sampling and overtone number.

Among the established methods for computing black-hole QNMs
\cite{Leaver:1985ax,Macedo:2016wgh,Guo:2022rms,Chandrasekhar:1975zza,Jansen:2017oag,Schutz:1985km,Konoplya:2003ii},
we employ a Chebyshev pseudospectral method to solve the coupled
system \cite{Jansen:2017oag}.
We validate the implementation by recovering the RN spectrum and
checking convergence with spectral order. We then determine how
the fundamental frequencies depend on the Rastall coupling, the
black-hole charge, and the strength of the surrounding matter.
The Rastall coupling affects the spectrum differently in the two
coupling regions considered here. In the configurations examined, increasing
the charge generally raises the oscillation frequency and changes
the damping rate nonmonotonically, while stronger surrounding-matter
contributions produce longer-lived modes.
All computed fundamental axial modes have
$\operatorname{Im}\omega<0$, consistently with the analytic
mode-stability result.

The remainder of this paper is organized as follows.
In Sec.~\ref{Background and perturbation equations}, we introduce
the background, derive the compatible matter response and coupled
master equations, analyze algebraic decoupling and the physical
dipole, and establish axial mode stability.
Section~\ref{Numerical Method and Quasinormal Mode Results}
presents the pseudospectral method, its validation, and the dependence
of the fundamental spectrum on the background parameters.
Section~\ref{conclusion} summarizes the results.
Appendix~\ref{app:axial_constraints} gives the component equations
and the identity relating the gravitational and matter constraints.
\section{Background, axial perturbations, and mode stability}
\label{Background and perturbation equations}

\subsection{Rastall gravity and black hole solutions} \label{Rastall gravity}

In general relativity, the energy-momentum tensor obeys the standard
covariant conservation law,
\begin{equation}
\nabla_{\mu}T^{\mu}{}_{\nu}=0.
\end{equation}
Rastall gravity~\cite{Rastall:1972swe} generalizes this condition by
allowing the divergence of the energy-momentum tensor to be
proportional to the gradient of the Ricci scalar,
\begin{equation}
\nabla_{\mu}T^{\mu}{}_{\nu}
=
\lambda \nabla_{\nu}R,
\label{eq:Rastall_nonconservation}
\end{equation}
where $\lambda$ is the Rastall parameter. The corresponding
gravitational field equations are
\begin{equation}
G_{\mu\nu}
+\kappa\lambda g_{\mu\nu}R
=
\kappa T_{\mu\nu},
\label{eq:Rastall_field_equation}
\end{equation}
where $\kappa$ denotes the gravitational coupling constant. The
standard Einstein equations are recovered in the limit
$\lambda\rightarrow0$, with $\kappa=8\pi G$ in conventional units.
We consider a static, spherically symmetric charged black hole
surrounded by a dust-like Kiselev-type anisotropic matter
distribution in Rastall gravity~\cite{Heydarzade:2017wxu}.
The gravitational field obeys the Rastall field equations,
while the electromagnetic field satisfies the source-free
Maxwell equations. Together with the surrounding matter,
these fields constitute the Einstein--Maxwell--Rastall system
considered here. We write the gravitational field equations as
\begin{equation}
G_{\mu\nu}+\kappa\lambda g_{\mu\nu}R
=
\kappa T_{\mu\nu}^{(\mathrm{EM})}
+\kappa T_{\mu\nu}^{(d)},
\label{eq:source_decomposition_background}
\end{equation}
where $T_{\mu\nu}^{(\mathrm{EM})}$ and $T_{\mu\nu}^{(d)}$
denote the electromagnetic and surrounding-matter
energy-momentum tensors, respectively.

In the Kiselev description, the equation-of-state parameter
characterizes the average pressure,
\[
\bar p_d\equiv\frac{p_r+2p_t}{3}
=\omega_d\rho_d,
\]
where $\rho_d$ is the surrounding-matter energy density,
and $p_r$ and $p_t$ are the radial and tangential pressures.
For the dust-like case $\omega_d=0$, the Kiselev stress tensor
satisfies
\[
p_r=-\rho_d,
\qquad
p_t=\frac{\rho_d}{2},
\]
so that $\bar p_d=0$ even though the individual pressures
are nonzero when $\rho_d\neq0$.
Thus, the term ``dust-like'' refers to the vanishing average
pressure; the source remains anisotropic and is distinct
from an isotropic pressureless fluid
\cite{Kiselev:2002dx,Visser:2019brz}.

The corresponding black-hole solution obtained in
Ref.~\cite{Heydarzade:2017wxu} has the line element
\begin{equation}
ds^{2}
=
-f(r)dt^{2}
+\frac{dr^{2}}{f(r)}
+r^{2}\left(d\theta^{2}+\sin^{2}\theta\,d\phi^{2}\right),
\label{eq:background_metric}
\end{equation}
with
\begin{equation}
f(r)
=
1-\frac{2M}{r}
+\frac{e^{2}}{r^{2}}
-\frac{N_d}{
r^{\frac{1-6\kappa\lambda}{1-3\kappa\lambda}}
}.
\label{eq:dust_metric_function}
\end{equation}
Here, $M$ is the mass parameter entering the metric,
$e$ is the electric-charge parameter, and $N_d$ is the
structure parameter controlling the contribution of the
surrounding matter to the background geometry.
For later convenience, we define the exponent
\begin{equation}
    \qquad
    \alpha\equiv\frac{1-6\kappa \lambda}{1-3\kappa\lambda}.
\end{equation}

Following Ref.~\cite{Heydarzade:2017wxu}, we impose the weak energy condition on
the surrounding matter and the strong energy condition on the
corresponding effective source. The parameter regions considered
in this work are
\begin{equation}
    0\leq\kappa\lambda<\frac{1}{6},
    \qquad N_d>0,
\end{equation}
and
\begin{equation}
    \kappa\lambda>\frac{1}{3},
    \qquad N_d<0.
\end{equation}
The intermediate interval $1/6<\kappa\lambda<1/3$
has $\alpha<0$, so the surrounding-matter contribution
does not decay at spatial infinity.
We therefore exclude this interval from the present analysis.

In both parameter regions considered here, $\alpha>0$.
Consequently, the surrounding-matter term vanishes
at large radius, and $f(r)\to1$ as $r\to\infty$.
The metric thus approaches the Minkowski form.
As shown explicitly in
Sec.~\ref{Coupled master equations},
the effective-potential matrix governing the axial
perturbations also vanishes at spatial infinity.
The asymptotic radial solutions therefore have
ingoing and outgoing wave factors
$e^{-i\omega r_*}$ and $e^{+i\omega r_*}$,
respectively, where $r_*$ is the tortoise coordinate.
For the time dependence $e^{-i\omega t}$,
the quasinormal-mode boundary condition selects
the purely outgoing branch at spatial infinity.

For $0<\kappa\lambda<1/6$, however, one has
$0<\alpha<1$, so the surrounding-matter term
falls off more slowly than the standard ADM
$1/r$ behavior.
In this region, $M$ is therefore interpreted as
the mass scale entering the metric rather than
as a finite ADM mass.

The horizons of the spacetime are determined by the positive real
roots of $f(r)=0$. We denote the largest positive root by $r_+$,
which corresponds to the outer event horizon. For a nonextremal
black hole with two positive roots, the other positive root is the
inner horizon $r_-$, with $r_-<r_+$.
In the extremal limit, the two horizons coalesce at
$r=r_{\rm ext}$, so that
\begin{equation}
    f(r_{\rm ext})=0,
    \qquad
    f'(r_{\rm ext})=0.
\end{equation}
For the metric function in Eq.~\eqref{eq:dust_metric_function}, these
conditions give
\begin{equation}
    r_{\rm ext}-M
    -\frac{2-\alpha}{2}N_d r_{\rm ext}^{\,1-\alpha}=0,
    \qquad
    e_{\rm ext}^{\,2}
    =M r_{\rm ext}
    +\frac{\alpha}{2}N_d r_{\rm ext}^{\,2-\alpha}.
\end{equation}
Thus, for fixed $(M,\kappa\lambda,N_d)$, the nonextremal black-hole
configurations considered below satisfy $|e|<e_{\rm ext}$, whereas
$|e|=e_{\rm ext}$ corresponds to the extremal limit. In the
Reissner--Nordstr\"om limit $N_d=0$, the above relations reduce to
$r_{\rm ext}=M$ and $|e_{\rm ext}|=M$.

\subsection{Axial perturbations} \label{Axial perturbations}

Owing to the spherical symmetry of the background spacetime, linear
perturbations can be decomposed into axial (odd-parity) and polar
(even-parity) sectors. Under the parity transformation
$(\theta,\phi)\rightarrow(\pi-\theta,\phi+\pi)$, the two sectors
transform as $(-1)^{\ell+1}$ and $(-1)^{\ell}$, respectively. In the
present work, we focus on the axial sector and adopt the
Regge--Wheeler gauge~\cite{Regge:1957td}.

We introduce first-order perturbations of the metric and the electromagnetic
potential as
\begin{equation}
g_{\mu\nu}
=
\bar{g}_{\mu\nu}
+\varepsilon h_{\mu\nu},
\qquad
A_{\mu}
=
\bar{A}_{\mu}
+\varepsilon a_{\mu},
\label{eq:linear_perturbations}
\end{equation}
where an overbar denotes a background quantity and $\varepsilon$ is a
bookkeeping parameter.
The source perturbations are treated according to the normalization
specified in Eq.~\eqref{eq:source_decomposition_background}. For
compactness, we define the perturbation of the total gravitational
source as
\begin{equation}
\delta \mathcal{T}_{\mu\nu}
=
\kappa\delta T_{\mu\nu}^{(\mathrm{EM})}
+
\kappa\delta T_{\mu\nu}^{(d)} .
\label{eq:total_gravitational_source_perturbation}
\end{equation}
Terms of order $\mathcal{O}(\varepsilon^{2})$ and higher are neglected.
The corresponding linearized Rastall field equations are
\begin{equation}
\delta G_{\mu\nu}
+
\kappa\lambda
\left(
h_{\mu\nu}\bar R+\bar g_{\mu\nu}\delta R
\right)
=
\delta \mathcal{T}_{\mu\nu}.
\label{eq:linearized_rastall_equation}
\end{equation}

All perturbation variables are assumed to have
the harmonic time dependence
\(\mathrm{e}^{-\mathrm{i}\omega t}\).
Since the perturbation equations are independent
of the azimuthal number \(m\), we set \(m=0\)
without loss of generality and define
\begin{equation}
\mathcal{S}_{\ell}(\theta)
=
\sin\theta\,
\frac{dP_{\ell}(\cos\theta)}{d\theta},
\label{eq:axial_angular_function}
\end{equation}
where $P_{\ell}(\cos\theta)$ is the Legendre polynomial.
In the Regge--Wheeler gauge, the axial metric perturbation is
described by two radial functions $h_{0}(r)$ and $h_{1}(r)$. In the
coordinate basis $(t,r,\theta,\phi)$, it takes the form~\cite{Regge:1957td}
\begin{equation}
h_{\mu\nu}
=
\mathrm{e}^{-\mathrm{i}\omega t}\mathcal{S}_{\ell}(\theta)
\begin{pmatrix}
0 & 0 & 0 & h_{0}(r) \\
0 & 0 & 0 & h_{1}(r) \\
0 & 0 & 0 & 0 \\
h_{0}(r) & h_{1}(r) & 0 & 0
\end{pmatrix}.
\label{eq:axial_metric_perturbation}
\end{equation}

The background electromagnetic potential is chosen as
\begin{equation}
\bar A_{\mu}dx^{\mu}
=
-\frac{e}{r}\,dt,
\label{eq:background_electromagnetic_potential}
\end{equation}
which gives the background field strength whose only
nonvanishing component is
\begin{equation}
\bar F_{tr}
=
-\frac{e}{r^{2}}.
\label{eq:background_field_strength}
\end{equation}
The axial electromagnetic perturbation is described by a single
radial function $f_{M}(r)$ and is written as~\cite{Zerilli:1974ai}
\begin{equation}
a_{\mu}dx^{\mu}
=
-f_{M}(r)\mathcal{S}_{\ell}(\theta)
\mathrm{e}^{-\mathrm{i}\omega t}\,d\phi.
\label{eq:axial_electromagnetic_perturbation}
\end{equation}
The corresponding field-strength perturbation,
\begin{equation}
\delta F_{\mu\nu}
=
\partial_{\mu}a_{\nu}
-
\partial_{\nu}a_{\mu},
\label{eq:perturbed_field_strength_definition}
\end{equation}
has the nonvanishing components
\begin{align}
\delta F_{t\phi}
&=
\mathrm{i}\omega f_{M}(r)
\mathcal{S}_{\ell}(\theta)\mathrm{e}^{-\mathrm{i}\omega t},
\\
\delta F_{r\phi}
&=
-f_{M}'(r)
\mathcal{S}_{\ell}(\theta)\mathrm{e}^{-\mathrm{i}\omega t},
\\
\delta F_{\theta\phi}
&=
\ell(\ell+1)f_{M}(r)
\sin\theta\,P_{\ell}(\cos\theta)\mathrm{e}^{-\mathrm{i}\omega t}.
\label{eq:perturbed_field_strength_components}
\end{align}
The electromagnetic energy-momentum tensor is then given by
\begin{equation}
T_{\mu\nu}^{(\mathrm{EM})}
=
\frac{2}{\kappa}
\left(
F_{\mu\alpha}F_{\nu}{}^{\alpha}
-
\frac{1}{4}g_{\mu\nu}
F_{\alpha\beta}F^{\alpha\beta}
\right).
\label{eq:electromagnetic_energy_momentum}
\end{equation}
Linearizing this expression with respect to the metric and
electromagnetic perturbations, the nonvanishing axial components are
\begin{align}
\delta T_{t\phi}^{(\mathrm{EM})}
&=
\frac{e}{\kappa r^{4}}
\left[
e h_{0}(r)
-
2r^{2}f(r)f_{M}'(r)
\right]
\mathcal{S}_{\ell}(\theta)\mathrm{e}^{-\mathrm{i}\omega t},
\label{eq:perturbed_energy_momentum_tphi}
\\
\delta T_{r\phi}^{(\mathrm{EM})}
&=
\frac{e}{\kappa r^{4}f(r)}
\left[
e f(r)h_{1}(r)
+
2\mathrm{i}\omega r^{2}f_{M}(r)
\right]
\mathcal{S}_{\ell}(\theta)\mathrm{e}^{-\mathrm{i}\omega t}.
\label{eq:perturbed_energy_momentum_rphi}
\end{align}
Their symmetric counterparts are understood, while the remaining
axial components vanish.

We next derive the axial contribution from the surrounding
Kiselev-type anisotropic matter source. This source
can be represented by an effective anisotropic fluid~\cite{Chen:2019iuo,Konoplya:2024lch},
\begin{equation}
T_{\mu\nu}^{(d)}
=
(\rho_d+p_t)u_\mu u_\nu
+
p_t g_{\mu\nu}
+
(p_r-p_t)s_\mu s_\nu ,
\label{eq:anisotropic_fluid_energy_momentum}
\end{equation}
where $u_\mu$ is the fluid four-velocity and $s_\mu$ is the radial unit
vector specifying the anisotropy direction. For the static background,
we take
\begin{equation}
\bar u_\mu=(-\sqrt{f},0,0,0),
\qquad
\bar s_\mu=(0,1/\sqrt{f},0,0).
\label{eq:background_fluid_vectors}
\end{equation}
The first-order perturbation of Eq.~\eqref{eq:anisotropic_fluid_energy_momentum}
is
\begin{align}
\delta T_{\mu\nu}^{(d)}
={}&
(\delta\rho_d+\delta p_t)\bar u_\mu\bar u_\nu
+
(\rho_d+p_t)
\left(
\delta u_\mu \bar u_\nu+\bar u_\mu\delta u_\nu
\right)
+
\delta p_t \bar g_{\mu\nu}
+
p_t h_{\mu\nu}
\nonumber\\
&+
(\delta p_r-\delta p_t)\bar s_\mu\bar s_\nu
+
(p_r-p_t)
\left(
\delta s_\mu \bar s_\nu+\bar s_\mu\delta s_\nu
\right).
\label{eq:anisotropic_fluid_linear_perturbation}
\end{align}
Because \(\delta\rho_d\), \(\delta p_r\), and
\(\delta p_t\) are scalar perturbations on the angular
two-sphere \(S^2\), their spherical-harmonic
decomposition has polar parity.
They therefore do not contribute to the axial sector,
and
\begin{equation}
\delta\rho_d=\delta p_r=\delta p_t=0
\label{eq:scalar_fluid_axial_perturbations}
\end{equation}
in the axial sector. We parameterize the axial vector perturbations as
\begin{equation}
\delta u_\phi
=
-\mathrm{i}\omega U(r)\mathcal S_\ell(\theta)
\mathrm{e}^{-\mathrm{i}\omega t},
\qquad
\delta s_\phi
=
-\Sigma(r)\mathcal S_\ell(\theta)\mathrm{e}^{-\mathrm{i}\omega t}.
\label{eq:fluid_vector_axial_perturbations}
\end{equation}
\subsubsection{Axial matter compatibility}
\label{subsec:axial_matter_compatibility}

We first derive the condition that the matter response must satisfy,
without fixing either $U$ or $\Sigma$. Since the Maxwell field obeys
the source-free Maxwell equations, its energy-momentum tensor is
conserved separately. The surrounding matter therefore obeys
\begin{equation}
\nabla_\mu T^{(d)\mu}{}_{\nu}
=\lambda\nabla_\nu R.
\label{eq:rastall_nonconservation_fluid}
\end{equation}
For the axisymmetric axial perturbations used here, the $\nu=\phi$
component has no contribution from the scalar gradient on the
right-hand side:
\begin{equation}
\partial_\phi\delta R=0,
\qquad
\delta\left(\nabla_\mu T^{(d)\mu}{}_{\phi}\right)=0.
\label{eq:phi_component_rastall_nonconservation}
\end{equation}
Before choosing a closure relation, write
the covariant axial source components as
\begin{equation}
\delta T^{(d)}_{t\phi}=\tau_t(r)\mathcal S_\ell(\theta)
\mathrm e^{-\mathrm i\omega t},
\qquad
\delta T^{(d)}_{r\phi}=\tau_r(r)\mathcal S_\ell(\theta)
\mathrm e^{-\mathrm i\omega t},
\label{eq:axial_source_amplitudes}
\end{equation}
with $\delta T^{(d)}_{\theta\phi}=0$, as follows from the
anisotropic-fluid stress tensor in
Eq.~\eqref{eq:anisotropic_fluid_linear_perturbation}.
Raising the first index also perturbs the inverse metric. The mixed
components are
\begin{align}
\delta T^{(d)t}{}_{\phi}
&=-\frac{\tau_t-p_t h_0}{f}\mathcal S_\ell
\mathrm e^{-\mathrm i\omega t},
\nonumber\\
\delta T^{(d)r}{}_{\phi}
&=f(\tau_r-p_t h_1)\mathcal S_\ell
\mathrm e^{-\mathrm i\omega t}.
\label{eq:axial_mixed_source}
\end{align}
The mixed components thus include the metric-induced contributions
proportional to $p_t h_0$ and $p_t h_1$. Substitution into
Eq.~\eqref{eq:phi_component_rastall_nonconservation} gives
\begin{equation}
\mathcal C_m\equiv
\frac{\mathrm i\omega}{f}(\tau_t-p_t h_0)
+\frac{1}{r^2}\left[r^2 f(\tau_r-p_t h_1)\right]'=0.
\label{eq:axial_matter_constraint}
\end{equation}
Primes on radial functions denote $d/dr$.

Equation~\eqref{eq:axial_matter_constraint} is the axial matter
compatibility condition. It links the matter and metric response
and must hold for solutions of the coupled perturbation equations.

To express this condition in terms of the fluid variables, define
\begin{equation}
W(r)\equiv\rho_d+p_t=3p_t,
\qquad p_r-p_t=-W.
\label{eq:axial_fluid_weight}
\end{equation}
Using Eqs.~\eqref{eq:anisotropic_fluid_linear_perturbation}
and~\eqref{eq:fluid_vector_axial_perturbations}, the source amplitudes are
\begin{equation}
\tau_t=p_t h_0+\mathrm i\omega W\sqrt f\,U,
\qquad
\tau_r=p_t h_1+\frac{W}{\sqrt f}\Sigma.
\label{eq:general_axial_fluid_response}
\end{equation}
Substitution into Eq.~\eqref{eq:axial_matter_constraint} gives
\begin{equation}
-\frac{\omega^2 W}{\sqrt f}U
+\frac{1}{r^2}\left(r^2\sqrt f\,W\Sigma\right)'=0.
\label{eq:axial_fluid_compatibility}
\end{equation}
For the dust-like Kiselev field considered here~\cite{Heydarzade:2017wxu}, the combination
entering the field equations is
\begin{equation}
\kappa p_t(r)
=
\frac{
3N_d\kappa\lambda(1-4\kappa\lambda)
}{
2(1-3\kappa\lambda)^2
}
r^{\frac{1}{1-3\kappa\lambda}-4}.
\label{eq:kpt_dust_kiselev}
\end{equation}
This pressure profile satisfies
\begin{equation}
p_t'=-\frac{\alpha+2}{r}p_t,
\qquad
\alpha=\frac{1-6\kappa\lambda}{1-3\kappa\lambda}.
\label{eq:axial_pressure_power}
\end{equation}
Where $W\neq0$, Eq.~\eqref{eq:axial_fluid_compatibility} becomes
\begin{equation}
\omega^2 U
=f\Sigma'+\left(\frac{f'}{2}-\frac{\alpha f}{r}\right)\Sigma.
\label{eq:axial_fluid_compatibility_reduced}
\end{equation}
The background pressure relations and this compatibility condition
leave the axial response underdetermined, requiring a closure for
$U$ and $\Sigma$. For $W=0$, the source-free limits follow directly
from the unreduced equations. We consider dynamical modes with
$\omega\neq0$ in the master-equation reduction below.

\subsubsection{A compatible matter-response closure}
\label{subsec:compatible_matter_response}

Following Ref.~\cite{Konoplya:2024lch}, we close the axial matter
system by holding the covariant anisotropy-direction vector fixed:
\begin{equation}
\delta s_\mu=0.
\label{eq:delta_s_zero_assumption}
\end{equation}
This prescription supplies the independent closure input for the
axial matter response. Setting $\Sigma=0$ in
Eq.~\eqref{eq:axial_fluid_compatibility} gives $U=0$ for
$\omega\neq0$ and $W\neq0$, and hence
\begin{equation}
\delta u_\phi=0.
\label{eq:delta_u_phi_zero}
\end{equation}
Consequently, the axial perturbation of the surrounding matter is
\begin{align}
\delta T_{t\phi}^{(d)}
&=p_t(r)h_0(r)\mathcal S_\ell(\theta)\mathrm e^{-\mathrm i\omega t},
\label{eq:fluid_perturbed_energy_momentum_tphi}\\
\delta T_{r\phi}^{(d)}
&=p_t(r)h_1(r)\mathcal S_\ell(\theta)\mathrm e^{-\mathrm i\omega t},
\label{eq:fluid_perturbed_energy_momentum_rphi}\\
\delta T_{\theta\phi}^{(d)}&=0.
\label{eq:fluid_perturbed_energy_momentum_thetaphi}
\end{align}
The metric-induced terms $\tau_t=p_t h_0$ and $\tau_r=p_t h_1$
satisfy $\mathcal C_m=0$ identically. This response closes the
gravitational and electromagnetic perturbation equations studied
below.

For later convenience, we introduce the Regge--Wheeler radial
variable \(Q(r)\) through
\begin{equation}
Q(r)
=
\frac{f(r)}{r}h_{1}(r),
\qquad
h_{1}(r)
=
\frac{r}{f(r)}Q(r).
\label{eq:regge_wheeler_variable}
\end{equation}
\subsection{Coupled master equations} \label{Coupled master equations}

For $\ell\geq2$, the angular gravitational equation imposes
$\mathcal C_g\equiv\mathrm i\omega h_0+f(fh_1)'=0$.
The identity in Eq.~\eqref{eq:axial_constraint_identity} relates the
temporal gravitational equation to the radial equation and the
matter constraint. With $\mathcal C_m=0$, the radial and angular
equations imply the temporal equation, ensuring a consistent
reduction to the master system. The component equations are given
in Appendix~\ref{app:axial_constraints}.
For $\ell=1$, the angular equation vanishes identically; the radial
constraint determines the physical electromagnetic dipole, derived
in Sec.~\ref{subsec:physical_axial_dipole}.

For the coupled $\ell\geq2$ system, substituting the compatible source
response and eliminating $h_0(r)$ gives two second-order equations
for $Q(r)$ and $f_M(r)$. To cast them into a Schr\"odinger-like form,
we introduce the master variables
\begin{equation}
\Psi_1(r)=Q(r),
\qquad
\Psi_2(r)=\mathrm{i}\omega f_M(r).
\label{eq:master_variables}
\end{equation}
together with the tortoise coordinate
\begin{equation}
\frac{d r_{*}}{d r}=\frac{1}{f(r)}.
\label{eq:tortoise_coordinate}
\end{equation}
The perturbation equations can then be expressed in the matrix form
\begin{equation}
\frac{d^{2}\boldsymbol{\Psi}}{d r_{*}^{2}}
+
\left[
\omega^{2}\mathbf{I}_{2}
-
\mathbf{V}(r)
\right]
\boldsymbol{\Psi}
=0,
\qquad
\boldsymbol{\Psi}
=
\begin{pmatrix}
\Psi_{1}\\
\Psi_{2}
\end{pmatrix},
\label{eq:coupled_schrodinger_equation}
\end{equation}
where $\mathbf{I}_{2}$ denotes the two-dimensional identity matrix.
The effective-potential matrix is defined by
\begin{equation}
\mathbf{V}(r)
=
\begin{pmatrix}
V_{11}(r) & V_{12}(r)\\
V_{21}(r) & V_{22}(r)
\end{pmatrix}.
\label{eq:effective_potential_matrix}
\end{equation}
For compactness, we introduce the dimensionless Rastall coupling
\begin{equation}
\gamma \equiv \kappa\lambda .
\label{eq:gamma_definition}
\end{equation}
In the following discussion,
we refer to the interval \(0\leq \gamma<1/6\)
as the small-\(\gamma\) region,
and to the interval \(\gamma>1/3\)
as the large-\(\gamma\) region.
In terms of $\gamma$, the components of the effective-potential matrix
are given by
\begin{align}
V_{11}(r)
={}&
f(r)
\left[
\frac{\ell(\ell+1)}{r^{2}}
-\frac{6M}{r^{3}}
+\frac{4e^{2}}{r^{4}}
+\frac{3N_{d}(4\gamma-1)}
{(1-3\gamma)}
r^{\frac{1}{1-3\gamma}-4}
\right],
\label{eq:V11}
\\
V_{12}(r)
={}&
-\frac{4e f(r)}{r^{3}},
\label{eq:V12}
\\
V_{21}(r)
={}&
-e f(r)
\frac{\ell(\ell+1)-2}{r^{3}},
\label{eq:V21}
\\
V_{22}(r)
={}&
f(r)
\left[
\frac{\ell(\ell+1)}{r^{2}}
+\frac{4e^{2}}{r^{4}}
\right].
\label{eq:V22}
\end{align}
Here, $V_{11}$ and $V_{22}$ represent the diagonal gravitational and
electromagnetic potentials, respectively, while $V_{12}$ and $V_{21}$
describe the coupling between the two perturbation sectors. The
off-diagonal terms are proportional to the black-hole charge $e$ and
therefore vanish in the neutral limit. 

As a consistency check, we take the
Reissner--Nordstr\"om limit by setting
\begin{equation}
N_d=0,
\qquad
\gamma=0.
\label{eq:RN_limit_conditions}
\end{equation}
The metric function then reduces to
\begin{equation}
f(r)
\longrightarrow
f_{\mathrm{RN}}(r)
=
1-\frac{2M}{r}
+\frac{e^{2}}{r^{2}}.
\label{eq:RN_metric_function}
\end{equation}
Under the same conditions,
Eqs.~\eqref{eq:V11}--\eqref{eq:V22}
recover the standard source-free axial
gravitoelectromagnetic perturbation system
of the RN black hole
\cite{Zerilli:1974ai}.
Thus, the present master equations reproduce
the standard Einstein--Maxwell result when
the surrounding Kiselev field and the Rastall
modification are both removed.


\subsection{Algebraic decoupling properties}
\label{Algebraic decoupling properties}

We now examine whether the coupled axial system can be reduced to two
independent Schr\"odinger-type equations by an $r$-independent linear
transformation. Consider the transformation
\begin{equation}
\boldsymbol{\Psi}
=
\mathbf P\boldsymbol{\Phi},
\qquad
\boldsymbol{\Phi}
=
\begin{pmatrix}
\Phi_{+}\\
\Phi_{-}
\end{pmatrix},
\label{eq:constant_decoupling_transformation}
\end{equation}
where $\mathbf P$ is an invertible matrix independent of $r$. The
transformed equations take the form
\begin{equation}
\frac{d^{2}\boldsymbol{\Phi}}{dr_{*}^{2}}
+
\left[
\omega^{2}\mathbf I_{2}
-
\mathbf P^{-1}\mathbf V(r)\mathbf P
\right]
\boldsymbol{\Phi}
=0.
\label{eq:transformed_coupled_equation}
\end{equation}
An $r$-independent algebraic decoupling therefore exists if the family
of potential matrices $\mathbf V(r)$ can be simultaneously
diagonalized. Away from isolated degeneracies, this can be tested
through the pairwise commutation condition
\begin{equation}
\left[
\mathbf V(r_{1}),
\mathbf V(r_{2})
\right]
=0
\qquad
\text{for arbitrary }r_{1}\text{ and }r_{2}.
\label{eq:simultaneous_diagonalization_condition}
\end{equation}

Pointwise diagonalization of $\mathbf V(r)$ is not sufficient to
decouple the differential equations. Indeed, if a radially dependent
transformation
\begin{equation}
\boldsymbol{\Psi}
=
\mathbf P(r)\boldsymbol{\Phi}
\label{eq:radial_decoupling_transformation}
\end{equation}
is employed, the transformed equations become
\begin{align}
\frac{d^{2}\boldsymbol{\Phi}}{dr_{*}^{2}}
&+
2\mathbf P^{-1}
\frac{d\mathbf P}{dr_{*}}
\frac{d\boldsymbol{\Phi}}{dr_{*}}
\nonumber\\
&+
\left[
\omega^{2}\mathbf I_{2}
-
\mathbf P^{-1}\mathbf V\mathbf P
+
\mathbf P^{-1}
\frac{d^{2}\mathbf P}{dr_{*}^{2}}
\right]
\boldsymbol{\Phi}
=0.
\label{eq:radially_transformed_equation}
\end{align}
Even when $\mathbf P^{-1}\mathbf V\mathbf P$ is diagonal, the derivative
term generally couples the transformed equations. Our decoupling
analysis therefore concerns algebraic transformations independent
of $r$.

For convenience, we define
\begin{equation}
L_{\ell}\equiv\ell(\ell+1),
\qquad
K_{\ell}\equiv L_{\ell}-2,
\qquad
\sigma\equiv \frac{1}{1-3\gamma},
\qquad
q\equiv\sigma-1=\frac{3\gamma}{1-3\gamma},
\label{eq:decoupling_parameters}
\end{equation}
together with the common scalar part
\begin{equation}
V_{0}(r;f)
\equiv
f(r)
\left[
\frac{L_{\ell}}{r^{2}}
+
\frac{4e^{2}}{r^{4}}
\right].
\label{eq:common_scalar_potential_decoupling}
\end{equation}
We first examine the RN and general-relativistic dust-like limits,
then turn to generic Rastall backgrounds.

\paragraph*{Reissner--Nordstr\"om electrovacuum.}

For \(N_d=0\) and \(\gamma=0\), the potential matrix
obtained from
Eqs.~\eqref{eq:V11}--\eqref{eq:V22}
can be decomposed as
\begin{equation}
\mathbf V_{\mathrm{RN}}(r)
=
V_{0}(r;f_{\mathrm{RN}})\mathbf I_{2}
+
\frac{f_{\mathrm{RN}}(r)}{r^{3}}
\mathbf A_{\mathrm{RN}},
\label{eq:RN_matrix_decomposition}
\end{equation}
where
\begin{equation}
\mathbf A_{\mathrm{RN}}
=
\begin{pmatrix}
-6M & -4e\\
-eK_{\ell} & 0
\end{pmatrix}.
\label{eq:RN_constant_matrix}
\end{equation}
Since $\mathbf A_{\mathrm{RN}}$ is independent of $r$, it can be
diagonalized by a constant transformation. Its eigenvalues are
\begin{equation}
\eta _{\pm}
=
-3M
\pm
\sqrt{
9M^{2}
+
4e^{2}K_{\ell}
},
\label{eq:RN_matrix_eigenvalues}
\end{equation}
and the two decoupled effective potentials are
\begin{equation}
V_{\pm}^{(\mathrm{RN})}(r)
=
V_{0}(r;f_{\mathrm{RN}})
+
\frac{f_{\mathrm{RN}}(r)}{r^{3}}
\eta _{\pm}.
\label{eq:RN_decoupled_potentials}
\end{equation}
Thus, the axial gravitational and electromagnetic perturbations of the
RN black hole can be expressed in terms of two independent master
variables. This constant algebraic decoupling of the coupled
Einstein--Maxwell perturbation system was obtained by
Zerilli~\cite{Zerilli:1974ai}. In the present formulation, it follows
directly from the fact that the nontrivial part of the potential matrix
is proportional to a single constant matrix.

\paragraph*{General-relativistic dust-like limit.}

In the limit $\gamma=0$ with $N_d\neq0$, the metric function becomes
\begin{equation}
f(r)
=
1-\frac{2M_{\mathrm{eff}}}{r}
+\frac{e^{2}}{r^{2}},
\qquad
M_{\mathrm{eff}}
\equiv
M+\frac{N_d}{2}.
\label{eq:effective_mass_GR_dust}
\end{equation}
The potential matrix has the same structure as
Eq.~\eqref{eq:RN_matrix_decomposition}, with
$M\rightarrow M_{\mathrm{eff}}$:
\begin{equation}
\mathbf V_{\mathrm{GR},d}(r)
=
V_{0}(r;f)\mathbf I_{2}
+
\frac{f(r)}{r^{3}}
\begin{pmatrix}
-6M_{\mathrm{eff}} & -4e\\
-eK_{\ell} & 0
\end{pmatrix}.
\label{eq:GR_dust_matrix_decomposition}
\end{equation}
Its nontrivial matrix part is again independent of $r$. The system
therefore remains algebraically decouplable, with eigenvalues
\begin{equation}
\eta _{\pm}^{(\mathrm{GR},d)}
=
-3M_{\mathrm{eff}}
\pm
\sqrt{
9M_{\mathrm{eff}}^{2}
+
4e^{2}K_{\ell}
}.
\label{eq:GR_dust_matrix_eigenvalues}
\end{equation}
Consequently, the dust-like term in the general-relativistic limit is
exactly degenerate with a redefinition of the mass parameter and does
not introduce an independent radial mixing structure.

\paragraph*{Generic Rastall background.}

For the matter response derived in
Sec.~\ref{subsec:compatible_matter_response}, the nonvanishing
covariant axial components are
\begin{equation}
\delta T_{t\phi}^{(d)}
=
p_t h_{t\phi},
\qquad
\delta T_{r\phi}^{(d)}
=
p_t h_{r\phi}.
\label{eq:matter_response_matter_perturbation}
\end{equation}
Using Eqs.~\eqref{eq:V11}--\eqref{eq:V22}, the potential matrix can be
rewritten as
\begin{equation}
\mathbf V(r)
=
V_{0}(r;f)\mathbf I_{2}
+
\frac{f(r)}{r^{3}}
\mathbf A(r),
\label{eq:matter_response_matrix_decomposition}
\end{equation}
where
\begin{equation}
\mathbf A(r)
=
\begin{pmatrix}
\displaystyle
-6M
+
\frac{
3N_d(4\gamma-1)
}{
1-3\gamma
}
r^{q}
&
-4e
\\[3ex]
-eK_{\ell}
&
0
\end{pmatrix}.
\label{eq:matter_response_radial_matrix}
\end{equation}
For \(e\neq0\), and away from points where
\(V_{12}=0\), write an eigenvector
of the potential matrix as
\begin{equation}
\boldsymbol v(r)
=
\begin{pmatrix}
1\\
x(r)
\end{pmatrix}.
\end{equation}
The component ratio \(x(r)\) satisfies
\begin{equation}
x^{2}
+
R_{2}(r)x
-
R_{1}(r)
=
0,
\qquad
R_{1}(r)
\equiv
\frac{V_{21}(r)}
     {V_{12}(r)},
\qquad
R_{2}(r)
\equiv
\frac{
V_{11}(r)
-
V_{22}(r)
}{
V_{12}(r)
}.
\label{eq:matter_response_eigenvector_ratio_condition}
\end{equation}
Therefore, away from isolated degeneracies,
an \(r\)-independent eigenbasis requires both
\(R_1(r)\) and \(R_2(r)\) to be independent of \(r\).

For the matter-response potential matrix,
the first ratio is
\begin{equation}
R_{1}(r)
=
\frac{
V_{21}(r)
}{
V_{12}(r)
}
=
\frac{K_{\ell}}{4},
\label{eq:matter_response_offdiagonal_ratio}
\end{equation}
which is independent of \(r\).
The remaining ratio determining the eigenvectors is
\begin{equation}
R_{2}(r)
=
\frac{
V_{11}(r)
-
V_{22}(r)
}{
V_{12}(r)
}
=
\frac{1}{4e}
\left[
6M
-
\frac{
3N_d(4\gamma-1)
}{
1-3\gamma
}
r^{q}
\right].
\label{eq:matter_response_decoupling_ratio}
\end{equation}
For generic \(N_d\neq0\) and
\(\gamma\notin\{0,1/4\}\),
\(R_2(r)\) depends explicitly on \(r\).
Consequently, \(x(r)\), and hence the eigenvectors
of the potential matrix, vary with the radial coordinate.
The matrices at different radii therefore do not possess
a common eigenbasis, and the system cannot be reduced
to two independent equations through a constant
linear combination of the master variables.

The RN and general-relativistic dust-like limits admit constant
algebraic decoupling because their potential matrices reduce to a
scalar part plus a single constant matrix. Generic charged Rastall
backgrounds with nonzero surrounding matter introduce additional
radial structure that obstructs this decoupling. The limits $N_d=0$
and $\gamma=0$ recover the RN-type structure, while $e=0$ separates
the gravitational and electromagnetic sectors directly. The formal
value $\gamma=1/4$ also removes the additional matrix entries but
lies outside the parameter regions studied here.

\subsection{Physical axial dipole}
\label{subsec:physical_axial_dipole}

For \(\ell=1\),
\begin{equation}
L_{\ell}=2,
\qquad
K_{\ell}=L_{\ell}-2=0.
\label{eq:dipole_angular_parameters}
\end{equation}
The angular equation vanishes identically because
$\mathcal S_1=-\sin^2\theta$ satisfies
$\partial_\theta\mathcal S_1-2\cot\theta\,\mathcal S_1=0$.
The radial constraint determines the physical dipole. For the
matter response $\tau_r=p_t h_1$,
$P h_1-\kappa\tau_r=0$ in
Eq.~\eqref{eq:axial_radial_equation}; with $K_1=0$ and
$\omega\neq0$, this requires $\mathcal Z=0$. Here $P=\kappa p_t$,
and $\mathcal Z$ is defined in Eq.~\eqref{eq:axial_auxiliary_variables}.
Substitution into the Maxwell equation~\eqref{eq:axial_maxwell_equation} gives
the autonomous electromagnetic equation below. Consistently, the
coupling potential satisfies
\begin{equation}
V_{21}=0,
\label{eq:dipole_v21}
\end{equation}
and the electromagnetic master equation becomes autonomous:
\begin{equation}
\frac{d^{2}\Psi_2}{dr_{*}^{2}}
+
\left[
\omega^{2}
-
V_{22}(r)
\right]\Psi_2
=0,
\label{eq:dipole_em}
\end{equation}
where
\begin{equation}
V_{22}(r)
=
f(r)
\left(
\frac{2}{r^{2}}
+
\frac{4e^{2}}{r^{4}}
\right).
\label{eq:dipole_v22}
\end{equation}
Once \(\Psi_2\) is determined, a metric reconstruction satisfying
$\mathcal C_g=0$ can be represented by
\begin{equation}
\frac{d^{2}\Psi_1}{dr_{*}^{2}}
+
\left[
\omega^{2}
-
V_{11}(r)
\right]\Psi_1
=
V_{12}(r)\Psi_2.
\label{eq:dipole_metric}
\end{equation}
The dipolar system has an upper-triangular structure.
The electromagnetic dipole propagates independently and induces a
metric response through \(V_{12}\).
The constraint $\mathcal Z=0$ selects the physical metric
reconstruction from solutions of this triangular system.
The radiative degree of freedom is an electromagnetic-led axial
dipole with an induced metric response.

We next establish mode stability for the coupled $\ell\geq2$ sector
and the physical electromagnetic dipole.

\subsection{Axial mode stability}
\label{subsec:axial_mode_stability}

We establish the absence of exponentially growing axial modes for
the matter response specified in
Sec.~\ref{subsec:compatible_matter_response}.
Consider a nonextremal background with a regular outer event
horizon at $r=r_+$, $f'(r_+)>0$, and a smooth static exterior
satisfying $f(r)>0$ for $r>r_+$ and $f(r)\to1$ as
$r\to\infty$. These conditions hold for the nonextremal
backgrounds considered here. The potential matrix is given by
Eqs.~\eqref{eq:V11}--\eqref{eq:V22}.

For $\ell\geq 2$, let
$L_\ell=\ell(\ell+1)$ and $K_\ell=L_\ell-2>0$.
The gravitational diagonal potential can be written in the
geometric form
\begin{equation}
 V_{11}
 =
 f\left(
 \frac{K_\ell}{r^2}
 +\frac{2f}{r^2}
 -\frac{f'}{r}
 \right),
 \label{eq:stability_geometric_V11}
\end{equation}
where $f'=df/dr$. Substitution of
Eq.~\eqref{eq:dust_metric_function} reproduces
Eq.~\eqref{eq:V11}, using
$\alpha+2=3(1-4\gamma)/(1-3\gamma)$.

Although $\mathbf V$ is not symmetric in the original master
variables, its off-diagonal entries have a constant ratio.
We therefore introduce the invertible constant transformation
\begin{equation}
 \widehat{\boldsymbol{\Psi}}
 =
 \mathbf T_\ell\boldsymbol{\Psi},
 \qquad
 \mathbf T_\ell
 =
 \operatorname{diag}\left(\frac{\sqrt{K_\ell}}{2},1\right).
 \label{eq:stability_symmetrization}
\end{equation}
Writing $x=r_*$, the master equations become
\begin{equation}
 -\frac{d^2\widehat{\boldsymbol{\Psi}}}{dx^2}
 +\widehat{\mathbf V}\widehat{\boldsymbol{\Psi}}
 =
 \omega^2\widehat{\boldsymbol{\Psi}},
 \qquad
 \widehat{\mathbf V}
 =
 \mathbf T_\ell\mathbf V\mathbf T_\ell^{-1}.
 \label{eq:stability_symmetric_system}
\end{equation}
The diagonal entries are unchanged, while
\[
 \widehat V_{12}
 =
 \widehat V_{21}
 =
 -\frac{2e\sqrt{K_\ell}f}{r^3}.
\]
The transformed potential $\widehat{\mathbf V}$ is real symmetric
and retains the coupling between the two channels.

We apply the matrix $S$-deformation method
\cite{Kimura:2018whv} with
\begin{equation}
 \mathbf S
 =
 \operatorname{diag}\left(\frac{f}{r},0\right),
 \qquad
 \mathcal D
 =
 \frac{d}{dx}+\mathbf S.
 \label{eq:stability_S}
\end{equation}
Since $d/dx=f\,d/dr$, the deformed potential is
\begin{equation}
 \begin{split}
 \widetilde{\mathbf V}
 &\equiv
 \widehat{\mathbf V}
 +\frac{d\mathbf S}{dx}
 -\mathbf S^2
 \\[2pt]
 &=
 \frac{f}{r^2}
 \begin{pmatrix}
 K_\ell
 &
 -2e\sqrt{K_\ell}/r
 \\[2pt]
 -2e\sqrt{K_\ell}/r
 &
 L_\ell+4e^2/r^2
 \end{pmatrix}.
 \end{split}
 \label{eq:stability_deformed_matrix}
\end{equation}
For any complex vector $\mathbf v=(v_1,v_2)^T$,
\begin{equation}
 \mathbf v^\dagger\widetilde{\mathbf V}\mathbf v
 =
 \frac{f}{r^2}
 \left[
 \left|
 \sqrt{K_\ell}v_1-\frac{2e}{r}v_2
 \right|^2
 +L_\ell|v_2|^2
 \right].
 \label{eq:stability_positive_form}
\end{equation}
Consequently, $\widetilde{\mathbf V}$ is positive definite
at every point of the static exterior. Equivalently, the
matrix in brackets in Eq.~\eqref{eq:stability_deformed_matrix}
has positive leading principal minor $K_\ell$ and determinant
$K_\ell L_\ell$. The $S$-deformation provides a positive
representation of the original radial operator's quadratic form.

To exclude unstable modes, suppose that a nontrivial solution
exists with $\operatorname{Im}\omega>0$ and with the
horizon-ingoing and infinity-outgoing boundary conditions.
Its radial dependence at the two endpoints is
\[
 \widehat{\boldsymbol{\Psi}}
 \sim
 \begin{cases}
 e^{-i\omega x}\mathbf a_H,
 & x\to-\infty,
 \\[2pt]
 e^{+i\omega x}\mathbf a_\infty,
 & x\to+\infty,
 \end{cases}
\]
where $\mathbf a_H$ and $\mathbf a_\infty$ are constant
amplitude vectors. For $\operatorname{Im}\omega>0$, these
solutions decay exponentially at their respective endpoints.
Moreover, $\mathbf S$ is bounded in the exterior and tends
to zero at both endpoints. The solution is therefore
square-integrable, and
\[
 \left[
 \widehat{\boldsymbol{\Psi}}^\dagger
 \mathcal D\widehat{\boldsymbol{\Psi}}
 \right]_{-\infty}^{+\infty}
 =0.
\]
Multiplying Eq.~\eqref{eq:stability_symmetric_system}
by $\widehat{\boldsymbol{\Psi}}^\dagger$, integrating,
and using this boundary behavior gives
\begin{equation}
 \begin{split}
 \omega^2
 \int_{-\infty}^{+\infty}
 \left|\widehat{\boldsymbol{\Psi}}\right|^2\,dx
 =
 \int_{-\infty}^{+\infty}
 \biggl[
 &
 \left|\mathcal D\widehat{\boldsymbol{\Psi}}\right|^2
 \\
 &+
 \widehat{\boldsymbol{\Psi}}^\dagger
 \widetilde{\mathbf V}
 \widehat{\boldsymbol{\Psi}}
 \biggr]\,dx .
 \end{split}
 \label{eq:stability_integral_identity}
\end{equation}
Here $|\mathbf v|^2=\mathbf v^\dagger\mathbf v$.
The right-hand side is real and nonnegative, so
$\omega^2$ must be real and nonnegative.
Writing $\omega=\omega_R+i\omega_I$ with $\omega_I>0$,
its imaginary part requires $2\omega_R\omega_I=0$,
and hence $\omega_R=0$. This would give
$\omega^2=-\omega_I^2<0$, a contradiction.
No exponentially growing coupled axial mode therefore exists
for $\ell\geq2$.

For $\ell=1$, $K_1=0$ and the transformation
in Eq.~\eqref{eq:stability_symmetrization} is singular,
so this case must be treated separately.
As derived from the radial constraint in
Sec.~\ref{subsec:physical_axial_dipole},
the physical electromagnetic dipole obeys
Eq.~\eqref{eq:dipole_em}, with
\[
 V_{\rm dipole}
 =
 f\left(\frac{2}{r^2}+\frac{4e^2}{r^4}\right)>0
 \qquad (r>r_+).
\]
The scalar version of the same integral argument, with
$S=0$, excludes exponentially growing dipolar modes.
The radial constraint ties the accompanying metric response
to this electromagnetic mode.

These results establish axial mode stability for every nonextremal
background satisfying the stated conditions and the adopted matter
closure. The proof covers all physical axial multipoles and
overtone numbers. Every hypothetical growing mode has $\omega\neq0$,
so the master-variable reduction remains applicable throughout
the argument.

\section{Numerical Method and Quasinormal Mode Spectrum} \label{Numerical Method and Quasinormal Mode Results}

\subsection{Pseudospectral formulation and boundary regularization}
\label{Pseudospectral Method}

We solve the coupled radial equations with the pseudospectral method~\cite{Jansen:2017oag}. Since the main numerical issue in the
present asymptotically Minkowskian problem is the nonstandard large-\(r\)
phase generated by the surrounding matter, we describe the boundary
regularization explicitly and keep the general discussion of spectral
interpolation brief.

For the numerical calculation, it is convenient to work with the
radial-amplitude vector
\begin{equation}
 \boldsymbol{\chi}(r)
 =
 \begin{pmatrix}
 Q(r)\\
 f_M(r)
 \end{pmatrix}.
 \label{eq:numerical_field_vector}
\end{equation}
This vector should be distinguished from
\(\boldsymbol{\Psi}\) in
Eq.~\eqref{eq:master_variables}: the two are related by
\(\boldsymbol{\Psi}=\operatorname{diag}(1,\mathrm{i}\omega)
\boldsymbol{\chi}\). For a nonzero quasinormal frequency, this
\(r\)-independent rescaling does not change the radial wave factors
or the boundary conditions of either component.

With the time dependence \(\mathrm{e}^{-\mathrm{i}\omega t}\), the
quasinormal boundary conditions are
\begin{equation}
 \boldsymbol{\chi}(r)
 \sim
 \begin{cases}
 \mathrm{e}^{-\mathrm{i}\omega r_*},
 & r\rightarrow r_+,
 \qquad \text{ingoing},\\[3pt]
 \mathrm{e}^{+\mathrm{i}\omega r_*},
 & r\rightarrow\infty,
 \qquad \text{outgoing},
 \end{cases}
 \label{eq:qnm_boundary_conditions}
\end{equation}
where \(r_*\) is defined in
Eq.~\eqref{eq:tortoise_coordinate}. We first extract the ingoing
horizon factor and write
\begin{equation}
 \boldsymbol{\chi}(r)
 =
 \mathrm{e}^{-\mathrm{i}\omega r_*(r)}
 \boldsymbol{\chi}_{\rm EF}(r).
 \label{eq:horizon_factorization}
\end{equation}
For a nonextremal horizon,
the tortoise coordinate behaves as
\[
r_*
\simeq
\frac{1}{f'(r_+)}
\ln(r-r_+).
\]
After extracting the ingoing factor in
Eq.~\eqref{eq:horizon_factorization},
the field
\(\boldsymbol{\chi}_{\rm EF}\)
is regular at \(r=r_+\).
Regularity of
\(\boldsymbol{\chi}_{\rm EF}\)
at the horizon therefore implements
the ingoing quasinormal-mode boundary condition.

We next introduce the compact coordinate
\begin{equation}
 u=\frac{r_+}{r},
 \qquad
 u\in[0,1],
 \label{eq:compact_coordinate}
\end{equation}
where \(u=0\) is spatial infinity and \(u=1\) is the event horizon.
The outgoing boundary condition at \(u=0\)
requires the large-\(r\) phase to be treated explicitly.
Because the surrounding-matter term may decay more slowly
than \(1/r\), the usual asymptotic factor containing only
\(r\) and \(\ln r\) is not sufficient.

To construct the outgoing phase,
we expand \(1/f(r)\) at large radius and retain
all terms that fall off no faster than \(1/r\):
\begin{equation}
\frac{1}{f(r)}
=
1+\frac{2M}{r}
+
\sum_{\substack{k\geq1\\ k\alpha\leq1}}
\frac{N_d^k}{r^{k\alpha}}
+
O\left(r^{-1-\delta}\right),
\qquad
\delta>0.
\label{eq:inverse_metric_asymptotics}
\end{equation}
Integrating precisely these terms gives
the asymptotic tortoise coordinate retained
in the numerical boundary factor,
\begin{equation}
\begin{split}
r_*^{(\infty)}(r)
={}&
r+2M\ln r
+
\sum_{\substack{k\geq1\\ k\alpha<1}}
\frac{N_d^k}{1-k\alpha}
r^{1-k\alpha}
+
\sum_{\substack{k\geq1\\ k\alpha=1}}
N_d^k\ln r .
\end{split}
\label{eq:asymptotic_tortoise_coordinate}
\end{equation}
Thus,
\(r_*^{(\infty)}\)
contains every asymptotic contribution
that grows as a power of \(r\)
or logarithmically at spatial infinity.
In the
small-\(\gamma\) interval used below,
\(7/16\leq\alpha\leq1\), so at most the \(k=1\) and \(k=2\)
matter terms occur. In particular, the logarithmic coefficient is
\(2M+N_d\) at \(\gamma=0\) (\(\alpha=1\)) and
\(2M+N_d^2\) at \(\gamma=1/9\) (\(\alpha=1/2\)). For
\(1/9<\gamma\leq0.12\), both growing powers
\(N_d r^{1-\alpha}/(1-\alpha)\) and
\(N_d^2 r^{1-2\alpha}/(1-2\alpha)\) are retained. In the
large-\(\gamma\) interval \(0.35\leq\gamma\leq1\), one has
\(\alpha\geq5/2\), and the sums in
Eq.~\eqref{eq:asymptotic_tortoise_coordinate} are empty.

The field that is regularized at both endpoints is then defined by
\begin{equation}
 \boldsymbol{\chi}(r)
 =
 \mathrm{e}^{-\mathrm{i}\omega r_*(r)}
 \mathcal{F}_{\infty}(u)
 \widetilde{\boldsymbol{\chi}}(u),
 \label{eq:full_boundary_factorization}
\end{equation}
where
\begin{equation}
 \mathcal{F}_{\infty}(u)
 =
 \exp\left(\frac{2\mathrm{i}\omega r_+}{u}\right)
 u^{-4\mathrm{i}M\omega}
 \exp\left[2\mathrm{i}\omega\Delta r_*(u)\right],
 \label{eq:outgoing_boundary_factor}
\end{equation}
and
\begin{equation}
 \Delta r_*(u)
 =
 r_*^{(\infty)}
 \left(\frac{r_+}{u}\right)
 -
 \frac{r_+}{u}
 -
 2M\ln\left(\frac{r_+}{u}\right).
 \label{eq:asymptotic_tortoise_remainder}
\end{equation}
Up to an irrelevant
\(u\)-independent normalization,
\[
\mathcal{F}_{\infty}
=
\exp\left(
2\mathrm{i}\omega
r_*^{(\infty)}
\right).
\]
The two factors in
Eq.~\eqref{eq:full_boundary_factorization}
therefore combine at spatial infinity to give
\[
\boldsymbol{\chi}
\sim
\mathrm{e}^{+\mathrm{i}\omega r_*}.
\]
Requiring
\(\widetilde{\boldsymbol{\chi}}\)
to remain regular at \(u=0\)
thus implements the outgoing
quasinormal-mode boundary condition.

After substituting
Eq.~\eqref{eq:full_boundary_factorization}
into the coupled radial equations,
the coefficients of the resulting equations
generally contain fractional powers of \(u\).
For the parameter values sampled in this work,
\(\alpha\) is represented by a rational number
and written in lowest terms as
\[
\alpha=\frac{z}{s}.
\]
We then introduce
\begin{equation}
u=y^s,
\qquad
y\in[0,1],
\label{eq:fractional_power_coordinate}
\end{equation}
which converts the powers
\(u^{m/s}\)
into integer powers of \(y\).
This coordinate regularization uses rational values of $\alpha$
for the numerical sampling. Irrational values can be treated with
a controlled rational approximation or a basis adapted to
noninteger powers.
After applying the coordinate transformation
to the coupled equations and clearing
the common powers of \(y\),
both equations have finite and nontrivial
limits at \(y=0\) and \(y=1\).
The endpoint equations are obtained from
these limits, and no additional numerical
boundary conditions are imposed.

The regular functions
\(\widetilde{\boldsymbol{\chi}}(y)\) are then discretized on the
Chebyshev--Lobatto grid
\begin{equation}
 y_j
 =
 \frac{1}{2}
 \left[
 1-\cos\left(\frac{j\pi}{N}\right)
 \right],
 \qquad
 j=0,\ldots,N.
 \label{eq:chebyshev_lobatto_grid}
\end{equation}
Here \(N\) denotes the spectral order, so that the grid contains
\(N+1\) collocation points, including the endpoints \(y=0\) and
\(y=1\).
Only their values at the grid points are retained, and radial
derivatives are replaced by the corresponding differentiation
matrices. As emphasized in Ref.~\cite{Jansen:2017oag}, once the
physical solutions have been made regular and the unwanted branches
nonregular, collocation on a smooth polynomial basis implements the
boundary selection implicitly.

Using \((Q,f_M)\) rather than
\((\Psi_1,\Psi_2)\)
introduces a factor \(1/\omega\)
in the second radial equation.
For the nonzero QNMs considered here,
we multiply this equation by \(\omega\)
to remove that denominator.
This operation also introduces
an additional algebraic zero-frequency sector,
which is excluded when the numerical
eigenvalues are filtered.
The discretized coupled system then takes
the polynomial form
\begin{equation}
 \mathcal{M}(\omega)\mathbf{v}
 =
 \left(
 \mathbf{M}_0
 +\omega\mathbf{M}_1
 +\omega^2\mathbf{M}_2
 +\omega^3\mathbf{M}_3
 \right)
 \mathbf{v}
 =0,
 \label{eq:polynomial_eigenvalue_problem}
\end{equation}
where \(\mathbf{v}\) contains the grid values of
\(\widetilde Q\) and \(\widetilde f_M\). Following
Ref.~\cite{Jansen:2017oag}, this polynomial problem is converted by
a companion linearization into a generalized eigenvalue problem.
Candidate eigenvalues are retained only when they remain stable as
the spectral order is increased. In practice, the spectra at
\(N=40\) and \(N=60\) are matched, and representative modes are
further checked at \(N=80\), as shown below.

\subsection{Quasinormal Mode Frequency} \label{QNM}




For the numerical results presented below, we use the
mass parameter $M$ as the reference scale and set $M=1$
in the calculations. We report the frequencies and
parameters in the dimensionless combinations
$M\omega$, $e/M$, and $N_d/M^\alpha$, where
$\alpha=(1-6\gamma)/(1-3\gamma)$.

We use the eigenvalue problem constructed above
to compute the dimensionless fundamental quasinormal
frequencies \(M\omega\) for different values of
the dimensionless charge \(e/M\), the Rastall coupling
\(\gamma\), and the surrounding-matter parameter
\(N_d/M^\alpha\).

For each spectral branch,
the fundamental mode corresponds to
the overtone number \(n=0\).
These are the least-damped modes within the RN-connected branches
tracked here.

Because the axial gravitational and
electromagnetic perturbations are coupled,
the modes discussed below are not purely
gravitational or purely electromagnetic.
The labels ``GR-led'' and ``EM-led''
denote the dominant component of the
corresponding eigenfunction.
The assignments are checked independently
by continuously tracking the associated
fundamental modes from the
RN limit.

The absence of exponentially growing modes in this
system has been established analytically in
Sec.~\ref{subsec:axial_mode_stability}.
Here we determine the fundamental oscillation frequencies
and damping rates and examine their dependence on the
background parameters.
The real part of the quasinormal frequency,
\(\operatorname{Re}\omega\),
determines the oscillation frequency,
whereas
\(|\operatorname{Im}\omega|\)
determines the damping rate.

\subsubsection{Validation and Branch Identification}

Before presenting the parameter dependence of the QNM spectrum,
we first perform several numerical checks.
We verify the RN limit,
identify the electromagnetic-led and gravitational-led branches
of the coupled \(\ell=2\) spectrum,
and examine the convergence with respect to the spectral order \(N\),
corresponding to \(N+1\) Chebyshev--Lobatto collocation points.

As a check of the numerical setup,
we first consider the RN limit
by setting \(N_d=0\) and \(\gamma=0\).
In this limit,
the background reduces to the RN spacetime,
whose quasinormal spectrum has been studied in detail
by Leaver~\cite{Leaver:1990zz}.
Following Leaver's notation,
we label the electromagnetic-type branch by \(i=1\)
and the gravitational-type branch by \(i=2\).
Table~\ref{tab:rn-limit}
lists the fundamental frequencies (\(n=0\))
for several values of the dimensionless charge parameter \(e/M\).
The numerical values agree with those reported
by Leaver~\cite{Leaver:1990zz},
providing a direct validation of our
numerical implementation.

\begin{table*}[htb]
\begin{center}
\begin{tabular}{cccc}
\hline\hline
\(e/M\)
& \(\ell=2,\ i=2\)
& \(\ell=2,\ i=1\)
& \(\ell=1,\ i=1\) \\
\hline

0.0
& \(0.37367168-0.08896231\,\mathrm{i}\)
& \(0.45759551-0.09500442\,\mathrm{i}\)
& \(0.24826326-0.09248771\,\mathrm{i}\) \\

0.2
& \(0.37474443-0.08907479\,\mathrm{i}\)
& \(0.46296518-0.09537344\,\mathrm{i}\)
& \(0.25147517-0.09290233\,\mathrm{i}\) \\

0.4
& \(0.37843689-0.08939811\,\mathrm{i}\)
& \(0.47992596-0.09644209\,\mathrm{i}\)
& \(0.26192174-0.09415595\,\mathrm{i}\) \\

0.6
& \(0.38621747-0.08981367\,\mathrm{i}\)
& \(0.51201092-0.09801663\,\mathrm{i}\)
& \(0.28275662-0.09620395\,\mathrm{i}\) \\

0.8
& \(0.40121719-0.08964323\,\mathrm{i}\)
& \(0.57013023-0.09906906\,\mathrm{i}\)
& \(0.32349360-0.09827250\,\mathrm{i}\) \\

\hline\hline
\end{tabular}
\caption{
Dimensionless fundamental quasinormal frequencies \(M\omega\)
in the Reissner--Nordstr\"om limit.
Here \(N_d/M^\alpha=0\), \(M=1\), and \(n=0\).
Following Leaver's notation,
the \(i=1\) branch corresponds to the electromagnetic-type mode,
whereas the \(i=2\) branch corresponds to the gravitational-type mode.
}
\label{tab:rn-limit}
\end{center}
\end{table*}

For nonzero \(e\),
the electromagnetic and gravitational perturbations
are coupled through the radial equations.
Therefore,
the two branches are no longer purely electromagnetic
or purely gravitational.
Following Ref.~\cite{Gu:2025lyz},
we identify the dominant character of each mode
by comparing the relative weights of the two components
in the corresponding eigenfunction.
In our notation, \(Q\) and \(f_M\) denote the gravitational and
electromagnetic radial amplitudes, respectively, whereas the
discretized eigenvector contains their regularized counterparts
\(\widetilde Q\) and \(\widetilde f_M\).
For each mode,
we define
\[
\mathcal{B}
=
\frac{\|\widetilde{Q}\|}{\|\widetilde{f_M}\|},
\]
where the norm is evaluated on the pseudospectral grid.
Modes with larger \(\mathcal{B}\) are identified as GR-led modes,
whereas modes with smaller \(\mathcal{B}\) are identified as EM-led modes.
Representative results are listed in Table~\ref{tab:branch-identification}.
\begin{table*}[htb]
\begin{center}
\begin{tabular}{ccccc}
\hline\hline
\(\gamma\)
& \(M\omega_{\rm GR}\)
& \(\mathcal{B}_{\rm GR}\)
& \(M\omega_{\rm EM}\)
& \(\mathcal{B}_{\rm EM}\) \\
\hline

0.02
& \(0.358667-0.084664\,\mathrm{i}\)
& \(1.544\)
& \(0.453326-0.091244\,\mathrm{i}\)
& \(0.110\) \\

0.04
& \(0.357110-0.084122\,\mathrm{i}\)
& \(1.541\)
& \(0.451281-0.090647\,\mathrm{i}\)
& \(0.109\) \\

0.06
& \(0.355159-0.083439\,\mathrm{i}\)
& \(1.539\)
& \(0.448732-0.089899\,\mathrm{i}\)
& \(0.108\) \\

0.08
& \(0.352657-0.082569\,\mathrm{i}\)
& \(1.536\)
& \(0.445477-0.088937\,\mathrm{i}\)
& \(0.107\) \\
\hline
0.40
& \(0.378463-0.089304\,\mathrm{i}\)
& \(1.554\)
& \(0.479975-0.096340\,\mathrm{i}\)
& \(0.122\) \\

0.60
& \(0.380437-0.088897\,\mathrm{i}\)
& \(1.553\)
& \(0.482747-0.095888\,\mathrm{i}\)
& \(0.124\) \\

0.80
& \(0.381904-0.089039\,\mathrm{i}\)
& \(1.554\)
& \(0.484781-0.096044\,\mathrm{i}\)
& \(0.125\) \\

1.00
& \(0.382759-0.089180\,\mathrm{i}\)
& \(1.553\)
& \(0.485961-0.096199\,\mathrm{i}\)
& \(0.125\) \\

\hline\hline
\end{tabular}
\caption{
Branch identification of the fundamental \(\ell=2\)
quasinormal modes.
Here \(e/M=0.4\).
For the small-\(\gamma\) region,
\(\gamma=0.02\), \(0.04\), \(0.06\) and \(0.08\), 
we take \(N_d/M^\alpha=0.1\).
For the large-\(\gamma\) region,
\(\gamma=0.40\), \(0.60\), \(0.80\) and \(1.00\),
we take \(N_d/M^\alpha=-0.1\).
The branch indicator is defined as
\(\mathcal{B}=\frac{\|\widetilde{Q}\|}{\|\widetilde{f_M}\|}\),
where \(Q\) and \(f_M\) denote the gravitational
and electromagnetic components of the eigenfunction,
respectively.
The two branches are clearly separated by the hierarchy of
\(\mathcal{B}\).
}
\label{tab:branch-identification}
\end{center}
\end{table*}

In addition to the component-ratio criterion described above, we
also identify the two branches by continuously tracking the
quasinormal frequencies from the Reissner--Nordstr\"om limit. Starting
from $N_d=0$, where the gravitational-type and electromagnetic-type
branches are unambiguously distinguished, we gradually vary the
background parameters toward their target values and follow the
continuous evolution of each mode. The branch assignments obtained
from this RN-limit tracking procedure agree with those determined
by the relative weights of the gravitational and electromagnetic
components. This agreement provides an additional consistency check
for our identification of the GR-led and EM-led branches.

We next examine the convergence of the numerical results
with respect to the spectral grid order.
For this purpose,
we choose several representative values of the dimensionless Rastall coupling
from both parameter regions considered here.
In the small-\(\gamma\) region,
we take \(\gamma=0.02\) and \(\gamma=0.10\)
with \(N_d/M^\alpha=0.1\),
whereas in the large-\(\gamma\) region,
we take \(\gamma=0.40\) and \(\gamma=0.80\)
with \(N_d/M^\alpha=-0.1\).
The charge parameter is fixed as \(e/M=0.4\).
For each parameter set,
the fundamental \(\ell=2\) quasinormal frequencies
are calculated using different spectral grid orders.
The results are shown in Table~\ref{tab:convergence}.
As the spectral order increases, the frequencies of both the GR-led
and EM-led branches approach stable values. For the tested
configurations, the $N=60$ and $N=80$ results agree to at least six
decimal places.
\begin{table*}[htb]
\begin{center}
\begin{tabular}{ccccc}
\hline\hline
\(\gamma\)
& Branch
& \(N=40\)
& \(N=60\)
& \(N=80\) \\
\hline

0.02
& GR-led
& \(0.358667-0.0846643\,\mathrm{i}\)
& \(0.358670-0.0846639\,\mathrm{i}\)
& \(0.358670-0.0846639\,\mathrm{i}\) \\

0.02
& EM-led
& \(0.453326-0.0912444\,\mathrm{i}\)
& \(0.453325-0.0912428\,\mathrm{i}\)
& \(0.453325-0.0912427\,\mathrm{i}\) \\

0.10
& GR-led
& \(0.349355-0.0814186\,\mathrm{i}\)
& \(0.349355-0.0814186\,\mathrm{i}\)
& \(0.349355-0.0814186\,\mathrm{i}\) \\

0.10
& EM-led
& \(0.441192-0.0876683\,\mathrm{i}\)
& \(0.441192-0.0876683\,\mathrm{i}\)
& \(0.441192-0.0876683\,\mathrm{i}\) \\
\hline
0.40
& GR-led
& \(0.378463-0.0893040\,\mathrm{i}\)
& \(0.378463-0.0893040\,\mathrm{i}\)
& \(0.378463-0.0893041\,\mathrm{i}\) \\

0.40
& EM-led
& \(0.479974-0.0963402\,\mathrm{i}\)
& \(0.479974-0.0963402\,\mathrm{i}\)
& \(0.479974-0.0963403\,\mathrm{i}\) \\

0.80
& GR-led
& \(0.381904-0.0890392\,\mathrm{i}\)
& \(0.381904-0.0890392\,\mathrm{i}\)
& \(0.381904-0.0890392\,\mathrm{i}\) \\

0.80
& EM-led
& \(0.484781-0.0960442\,\mathrm{i}\)
& \(0.484781-0.0960442\,\mathrm{i}\)
& \(0.484781-0.0960442\,\mathrm{i}\) \\

\hline\hline
\end{tabular}
\caption{The tabulated frequencies are the dimensionless quantities \(M\omega\).
Convergence test of the fundamental \(\ell=2\)
quasinormal frequencies with respect to
the spectral grid order \(N\).
The corresponding number of Chebyshev--Lobatto collocation
points is \(N+1\).
The charge parameter is fixed as \(e/M=0.4\).
For the small-\(\gamma\) region,
we take \(N_d/M^\alpha=0.1\),
whereas for the large-\(\gamma\) region,
we take \(N_d/M^\alpha=-0.1\).
}
\label{tab:convergence}
\end{center}
\end{table*}

\subsubsection{Influence of the Dimensionless Rastall Coupling \(\gamma\)}

\begin{figure*}[htb]
    \begin{center}
    \subfigure[\hspace{0.4em}\(\ell=1\)]{\label{fig:qnm-lambda-small-l1}
    \includegraphics[width=15.0cm]{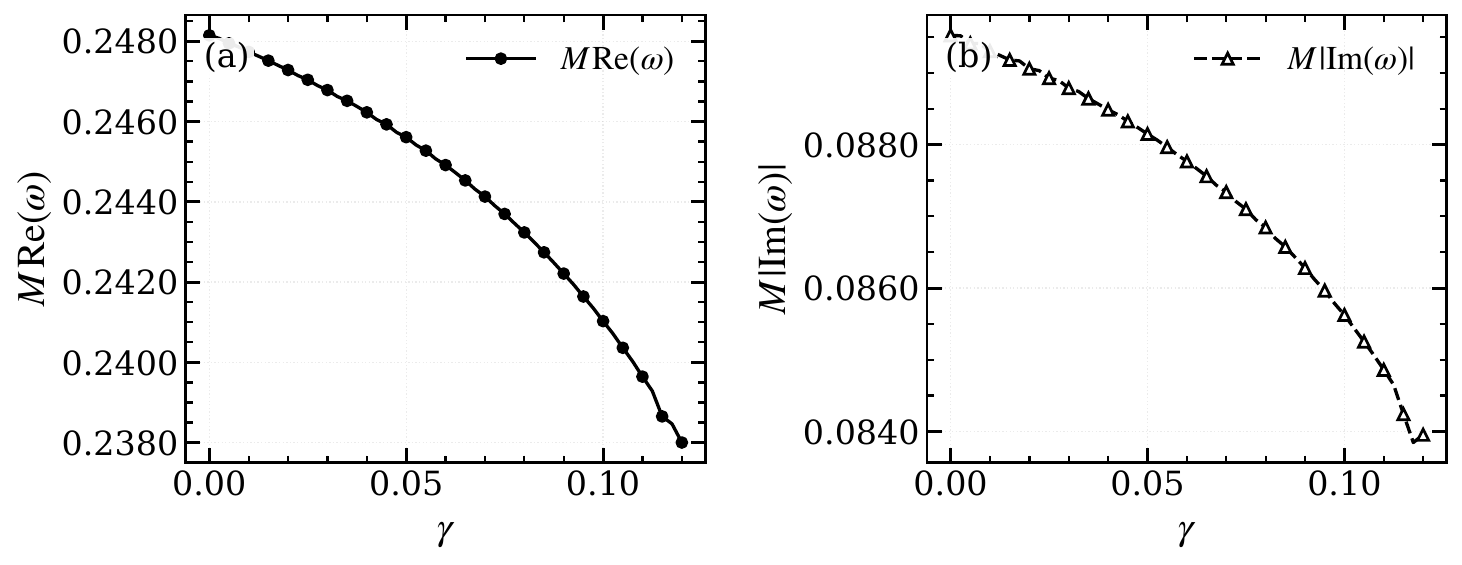}}

    \vspace{0.4cm}

    \subfigure[\hspace{0.4em}\(\ell=2\)]{\label{fig:qnm-lambda-small-l2}
    \includegraphics[width=15.0cm]{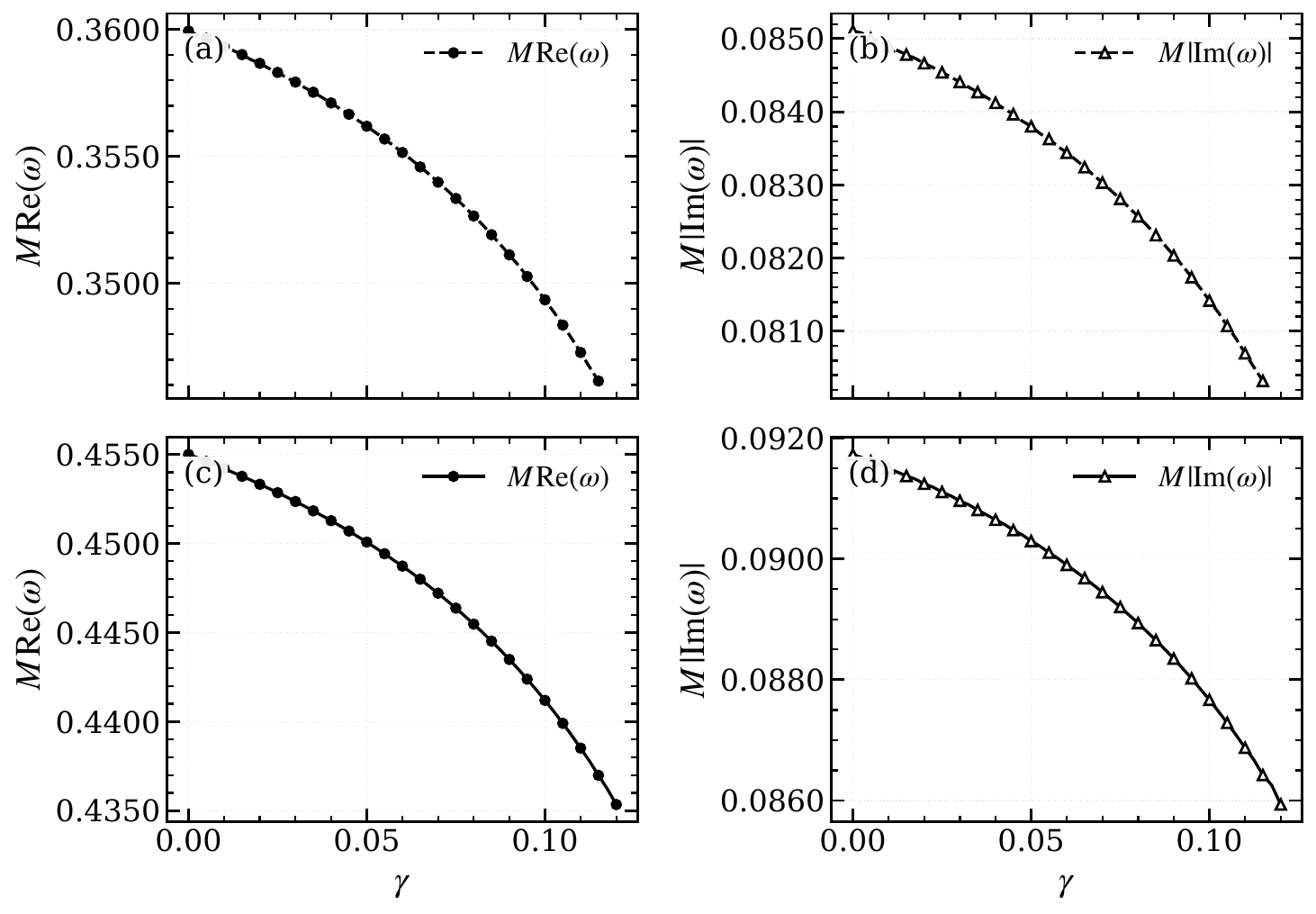}}
    \end{center}
    \caption{
    The dependence of the dimensionless fundamental
    quasinormal frequencies \(M\omega\)
    on the dimensionless Rastall coupling \(\gamma\)
    in the small-\(\gamma\) region.
    The parameters are fixed as \(e/M=0.4\) and \(N_d/M^\alpha=0.1\).
    For the \(\ell=2\) case,
    the upper two panels correspond to the GR-led branch,
    while the lower two panels correspond to the EM-led branch.
    }
    \label{fig:qnm-lambda-small}
\end{figure*}

\begin{figure*}[htb]
    \begin{center}
    \subfigure[\hspace{0.4em}\(\ell=1\)]{\label{fig:qnm-lambda-large-l1}
    \includegraphics[width=15.0cm]{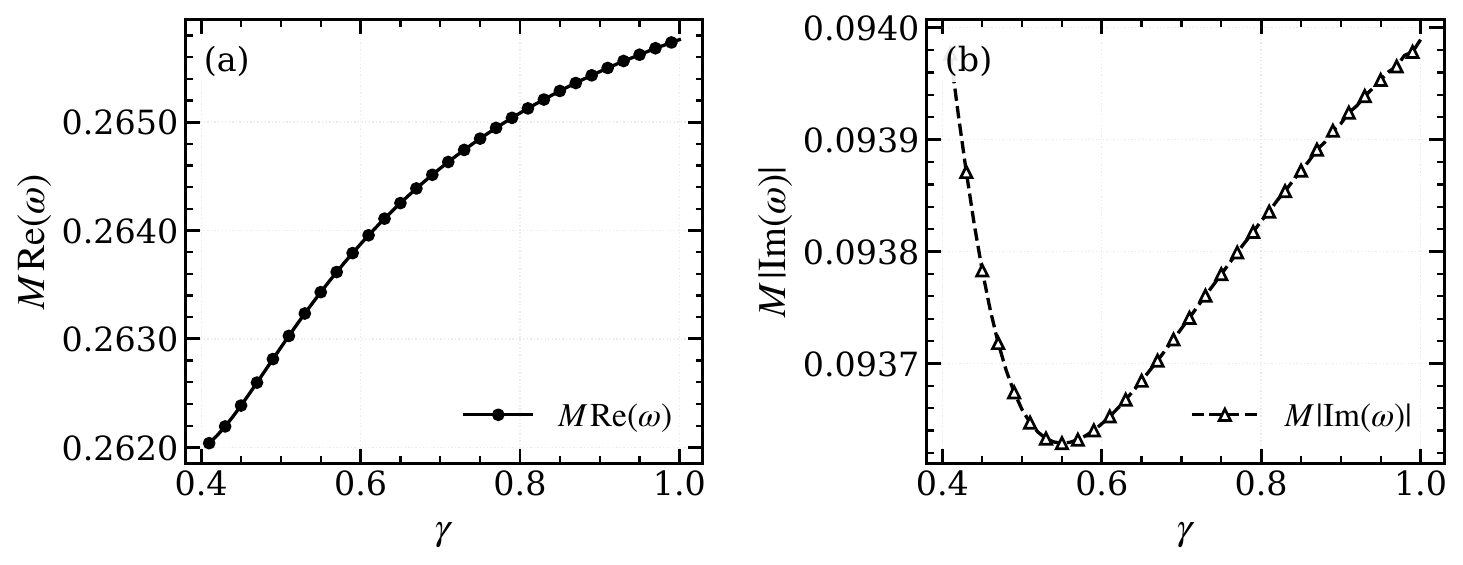}}

    \vspace{0.4cm}

    \subfigure[\hspace{0.4em}\(\ell=2\)]{\label{fig:qnm-lambda-large-l2}
    \includegraphics[width=15.0cm]{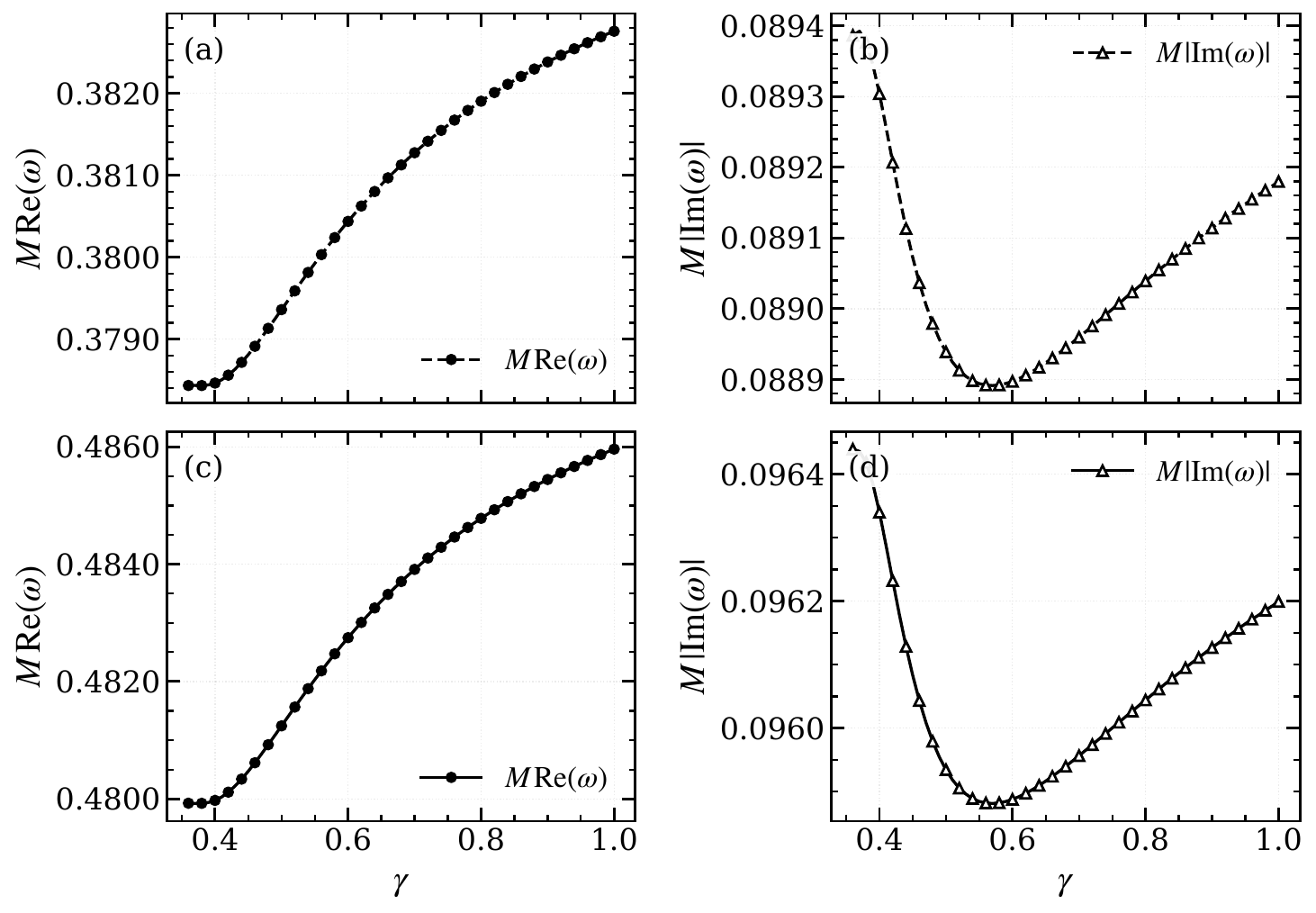}}
    \end{center}
    \caption{
    The dependence of the dimensionless fundamental
    quasinormal frequencies \(M\omega\)
    on the dimensionless Rastall coupling \(\gamma\)
    in the large-\(\gamma\) region.
    The parameters are fixed as \(e/M=0.4\) and \(N_d/M^\alpha=-0.1\).
    For the \(\ell=2\) case, the upper two panels correspond to the GR-led branch,
    while the lower two panels correspond to the EM-led branch.
    }
    \label{fig:qnm-lambda-large}
\end{figure*}

In this subsection,
we investigate the influence of the dimensionless Rastall coupling \(\gamma\)
on the fundamental quasinormal frequencies.
The numerical results are shown in
Figs.~\ref{fig:qnm-lambda-small}
and \ref{fig:qnm-lambda-large}.

The small-\(\gamma\) region includes the GR boundary at \(\gamma=0\)
and extends over \(0\leq\gamma<1/6\).
Within this region,
we restrict the small-\(\gamma\) numerical analysis to
\(0\leq\gamma\leq0.12\)
in order to stay away from the critical value \(\gamma=1/6\).
For the large-\(\gamma\) region,
where \(\gamma>1/3\),
we choose the representative interval
\(0.35\leq\gamma\leq1\).
At \(\gamma=1/6\), \(\alpha=0\) and the surrounding-matter term
becomes constant, whereas \(\alpha\) diverges as
\(\gamma\to(1/3)^+\).
Our numerical scans exclude these limiting regimes.
Representative spectral-convergence tests are presented in
Table~\ref{tab:convergence}.

In the small-\(\gamma\) region,
Fig.~\ref{fig:qnm-lambda-small}
shows that both
\(M\operatorname{Re}\omega\)
and
\(M|\operatorname{Im}\omega|\)
decrease monotonically as \(\gamma\) increases.
This behavior occurs for the \(\ell=1\) mode
and for both the GR-led and EM-led
\(\ell=2\) branches.
Thus, increasing the Rastall coupling
in this region lowers the oscillation frequency
and weakens the damping,
leading to longer-lived perturbations.

By contrast,
the real parts of the frequencies increase
with \(\gamma\) in the large-\(\gamma\) region,
as shown in
Fig.~\ref{fig:qnm-lambda-large}.
The damping rate is nonmonotonic:
\(M|\operatorname{Im}\omega|\)
first decreases and then increases,
with a minimum inside the parameter interval.
The perturbations therefore become
initially longer-lived and subsequently
more strongly damped as \(\gamma\) increases.

Taken together,
these results show that the spectral response
to the Rastall coupling depends on the
coupling region considered here.
For \(\ell=2\),
the EM-led branch has a higher oscillation
frequency than the GR-led branch,
while the two branches exhibit the same
qualitative dependence on \(\gamma\).

\subsubsection{Influence of the Dimensionless Charge Parameter \(e/M\)}

We now study the influence of the black-hole dimensionless charge parameter \(e/M\)
on the fundamental quasinormal frequencies.
In order to compare the two coupling regions considered here,
we again choose two representative cases.
For the small-\(\gamma\) region,
we fix \(\gamma=0.1\) and \(N_d/M^\alpha=0.1\),
whereas for the large-\(\gamma\) region,
we fix \(\gamma=0.5\) and \(N_d/M^\alpha=-0.1\).
For the two representative backgrounds considered below, the
extremal charges are $e_{\rm ext}/M\simeq1.05131$ for
$(\gamma,N_d/M^\alpha)=(0.1,0.1)$ and $e_{\rm ext}/M\simeq0.95282$ for
$(\gamma,N_d/M^\alpha)=(0.5,-0.1)$. The charge ranges used in Figs.~3 and 4
therefore remain within the nonextremal black-hole regime.

\begin{figure*}[htb]
    \begin{center}
    \subfigure[\hspace{0.4em}\(\ell=1\)]{\label{fig:qnm-e-small-l1}
    \includegraphics[width=15.0cm]{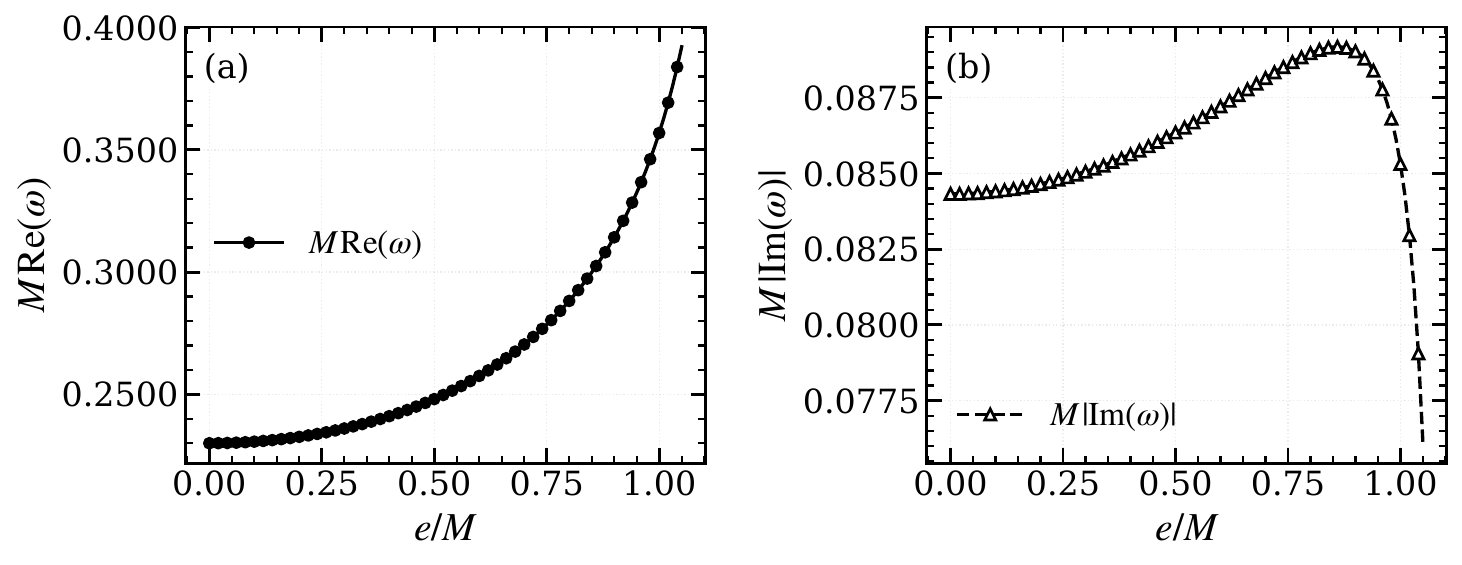}}

    \vspace{0.4cm}

    \subfigure[\hspace{0.4em}\(\ell=2\)]{\label{fig:qnm-e-small-l2}
    \includegraphics[width=15.0cm]{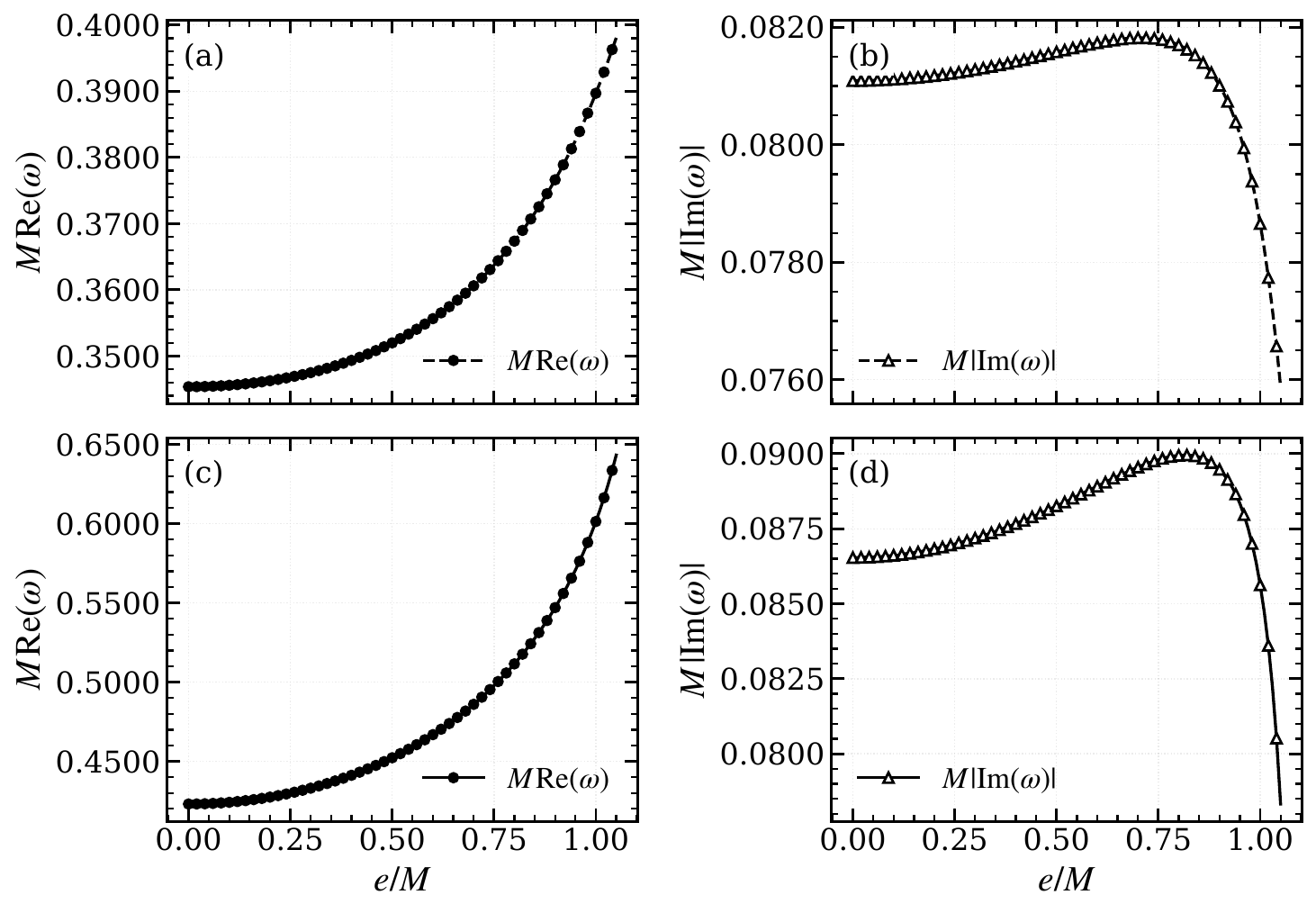}}
    \end{center}
    \caption{
    The dependence of the dimensionless fundamental
    quasinormal frequencies \(M\omega\)
    on the dimensionless charge parameter \(e/M\)
    in the small-\(\gamma\) region.
    The parameters are fixed as \(\gamma=0.1\) and \(N_d/M^\alpha=0.1\).
    For the \(\ell=2\) case,
    the upper two panels correspond to the GR-led branch,
    while the lower two panels correspond to the EM-led branch.
    }
    \label{fig:qnm-e-small}
\end{figure*}

For \(\gamma=0.1\) and \(N_d/M^\alpha=0.1\),
Fig.~\ref{fig:qnm-e-small}
shows that
\(M\operatorname{Re}\omega\)
increases monotonically with \(e/M\)
for the \(\ell=1\) mode
and for both \(\ell=2\) branches.
The EM-led branch remains at a higher
oscillation frequency than the GR-led branch.
The damping rate is nonmonotonic.
Starting from small charge,
\(M|\operatorname{Im}\omega|\)
increases to a maximum and then decreases
as \(e/M\) is increased further.
Thus, the perturbations initially decay faster,
whereas sufficiently highly charged
configurations support longer-lived modes.

\begin{figure*}[htb]
    \begin{center}
    \subfigure[\hspace{0.4em}\(\ell=1\)]{\label{fig:qnm-e-large-l1}
    \includegraphics[width=15.0cm]{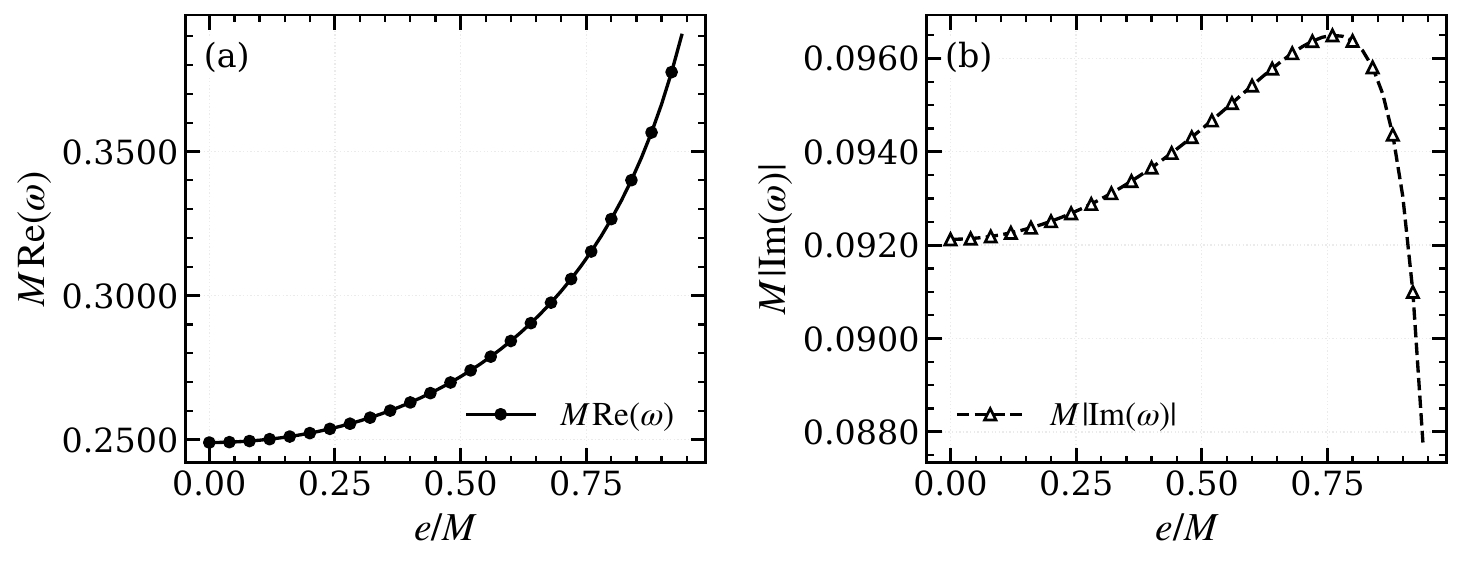}}

    \vspace{0.4cm}

    \subfigure[\hspace{0.4em}\(\ell=2\)]{\label{fig:qnm-e-large-l2}
    \includegraphics[width=15.0cm]{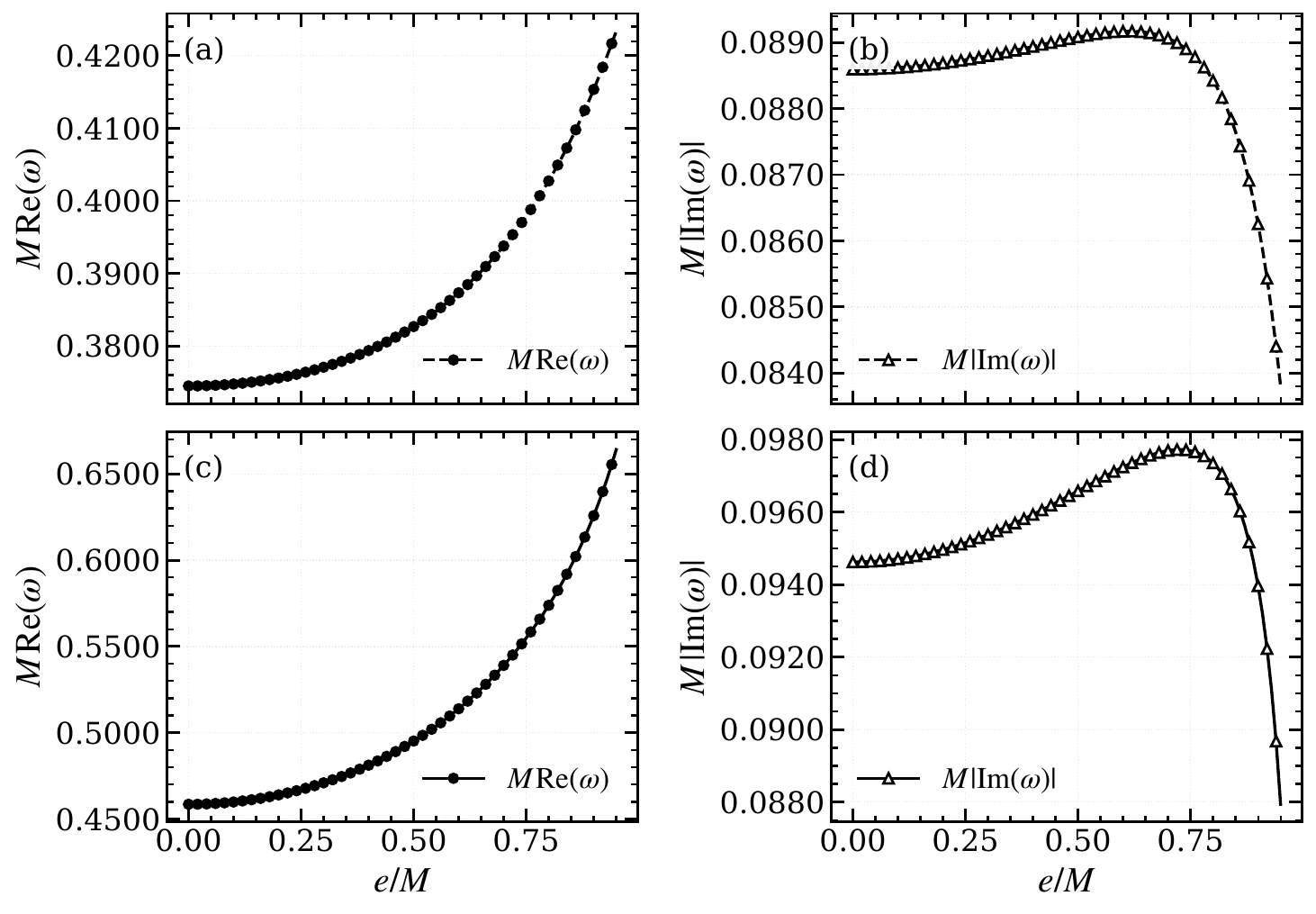}}
    \end{center}
    \caption{
    The dependence of the dimensionless fundamental
    quasinormal frequencies \(M\omega\)
    on the dimensionless charge parameter \(e/M\)
    in the large-\(\gamma\) region.
    The parameters are fixed as \(\gamma=0.5\) and \(N_d/M^\alpha=-0.1\).
    For the \(\ell=2\) case,
    the upper two panels correspond to the GR-led branch,
    while the lower two panels correspond to the EM-led branch.
    }
    \label{fig:qnm-e-large}
\end{figure*}


A similar charge dependence is found
in the large-\(\gamma\) region,
as shown in
Fig.~\ref{fig:qnm-e-large}.
The real parts of the frequencies increase
monotonically with \(e/M\),
whereas
\(M|\operatorname{Im}\omega|\)
first increases and then decreases.
This behavior is shared by the \(\ell=1\) mode
and by both \(\ell=2\) branches.

Therefore,
in both Rastall-coupling regions,
a larger black-hole charge raises
the oscillation frequency.
Its effect on the damping rate is nonmonotonic:
the damping strengthens at moderate charge
and weakens again at larger charge.
\subsubsection{Influence of the Dimensionless Surrounding-Matter Parameter \(N_d/M^\alpha\)}
We now investigate the influence of the dimensionless surrounding-matter parameter
\(N_d/M^\alpha\) on the fundamental quasinormal frequencies.
The charge parameter is fixed as \(e/M=0.4\).
In order to compare the two coupling regions considered here,
we consider two representative values,
\(\gamma=0.1\) and \(\gamma=0.5\).
According to the parameter constraints discussed above,
the former belongs to the small-\(\gamma\) region
and is associated with positive \(N_d/M^\alpha\),
whereas the latter belongs to the large-\(\gamma\) region
and is associated with negative \(N_d/M^\alpha\).

\begin{figure*}[htb]
    \begin{center}
    \subfigure[\hspace{0.4em}\(\ell=1\)]{\label{fig:qnm-Nd-small-l1}
    \includegraphics[width=15.0cm]{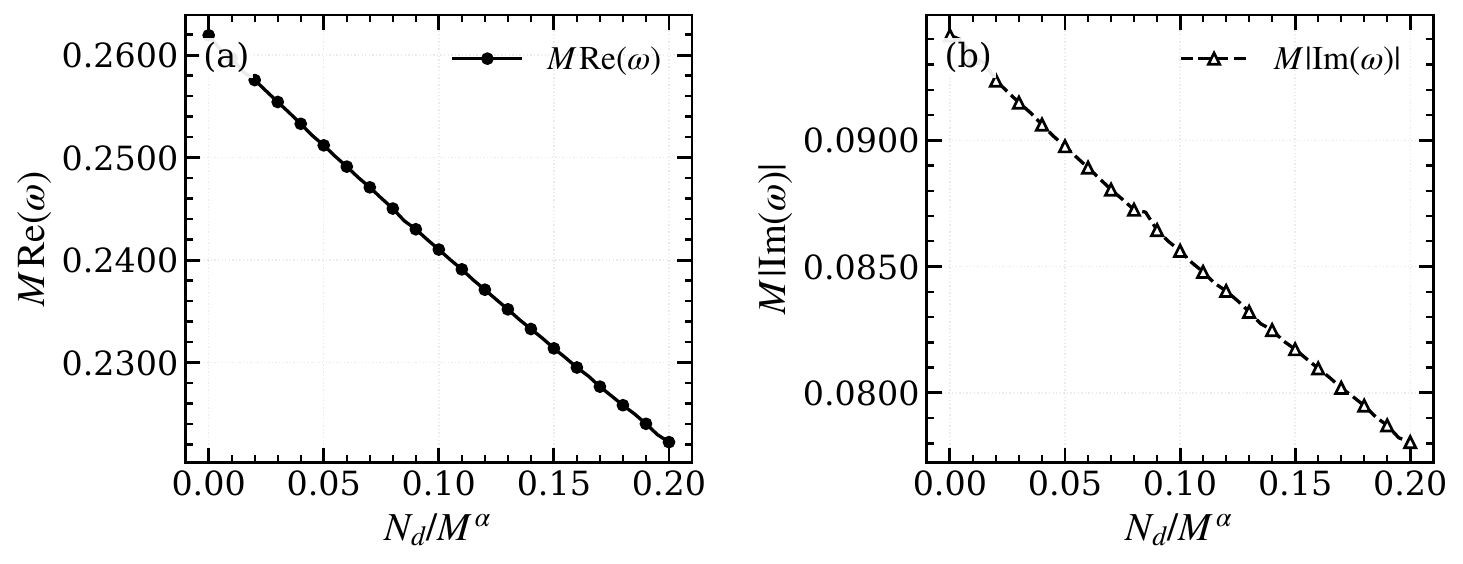}}

    \vspace{0.4cm}

    \subfigure[\hspace{0.4em}\(\ell=2\)]{\label{fig:qnm-Nd-small-l2}
    \includegraphics[width=15.0cm]{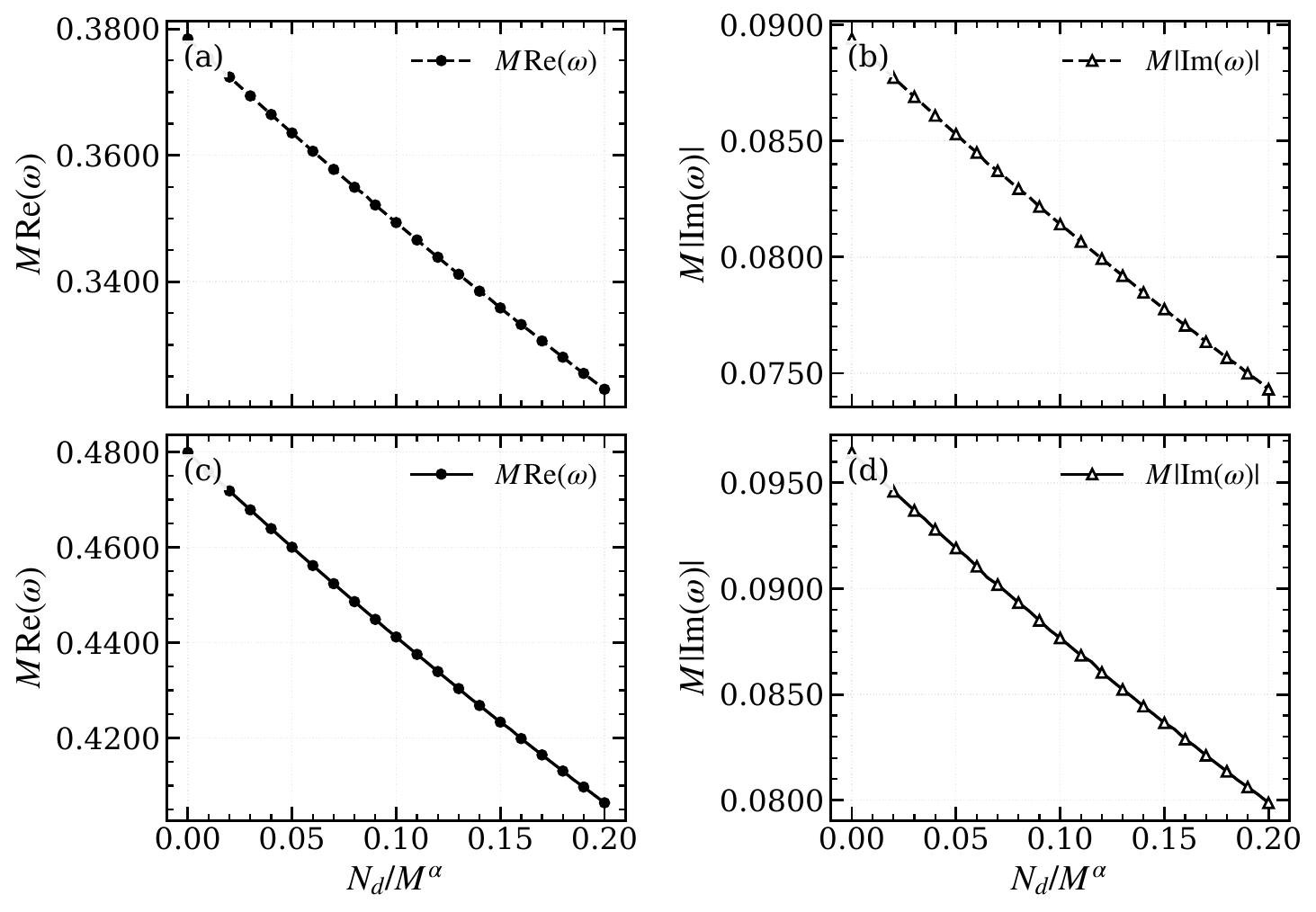}}
    \end{center}
    \caption{
    The dependence of the dimensionless fundamental
    quasinormal frequencies \(M\omega\)
    on the dimensionless surrounding-matter parameter \(N_d/M^\alpha\)
    in the small-\(\gamma\) region.
    The parameters are fixed as \(e/M=0.4\) and \(\gamma=0.1\).
    For the \(\ell=2\) case,
    the upper two panels correspond to the GR-led branch,
    while the lower two panels correspond to the EM-led branch.
    }
    \label{fig:qnm-Nd-small}
\end{figure*}

For \(\gamma=0.1\),
Fig.~\ref{fig:qnm-Nd-small}
shows that both
\(M\operatorname{Re}\omega\)
and
\(M|\operatorname{Im}\omega|\)
decrease monotonically as the positive
parameter \(N_d/M^\alpha\) increases.
This behavior occurs for the \(\ell=1\) mode
and for both \(\ell=2\) branches.
Increasing the surrounding-matter contribution
therefore lowers the oscillation frequency,
while the simultaneous decrease of
\(|M\operatorname{Im}\omega|\)
indicates weaker damping and longer-lived modes.

\begin{figure*}[htb]
    \begin{center}
    \subfigure[\hspace{0.4em}\(\ell=1\)]{\label{fig:qnm-Nd-large-l1}
    \includegraphics[width=15.0cm]{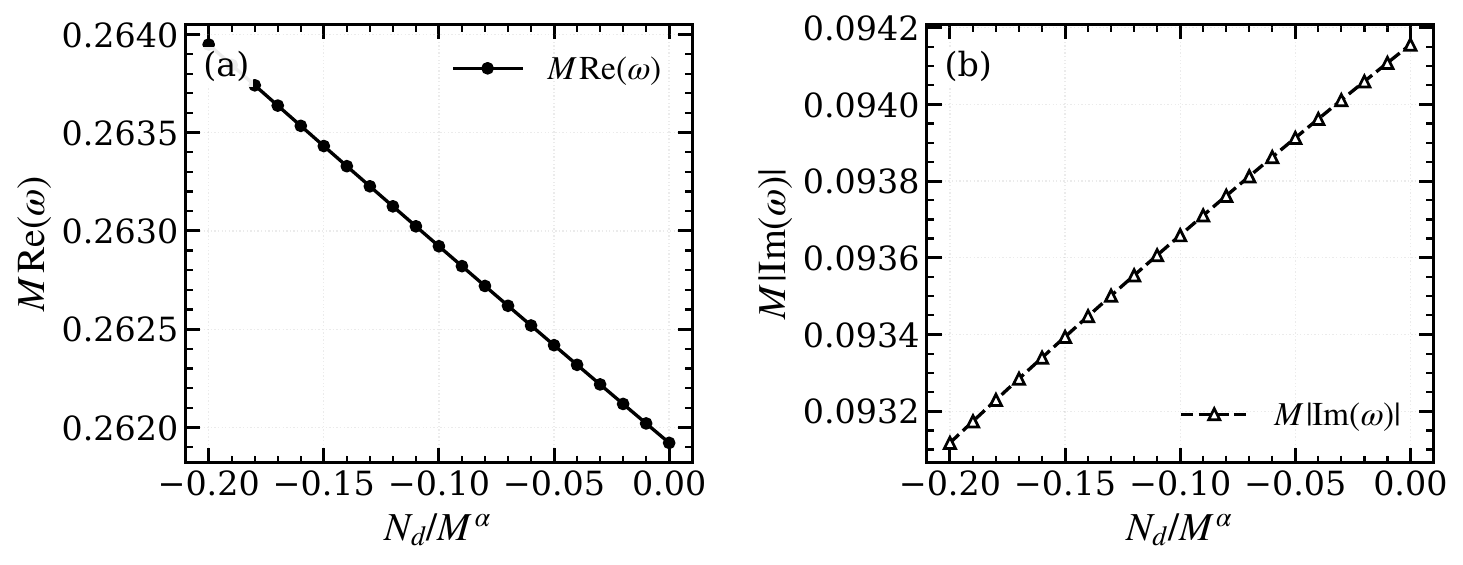}}

    \vspace{0.4cm}

    \subfigure[\hspace{0.4em}\(\ell=2\)]{\label{fig:qnm-Nd-large-l2}
    \includegraphics[width=15.0cm]{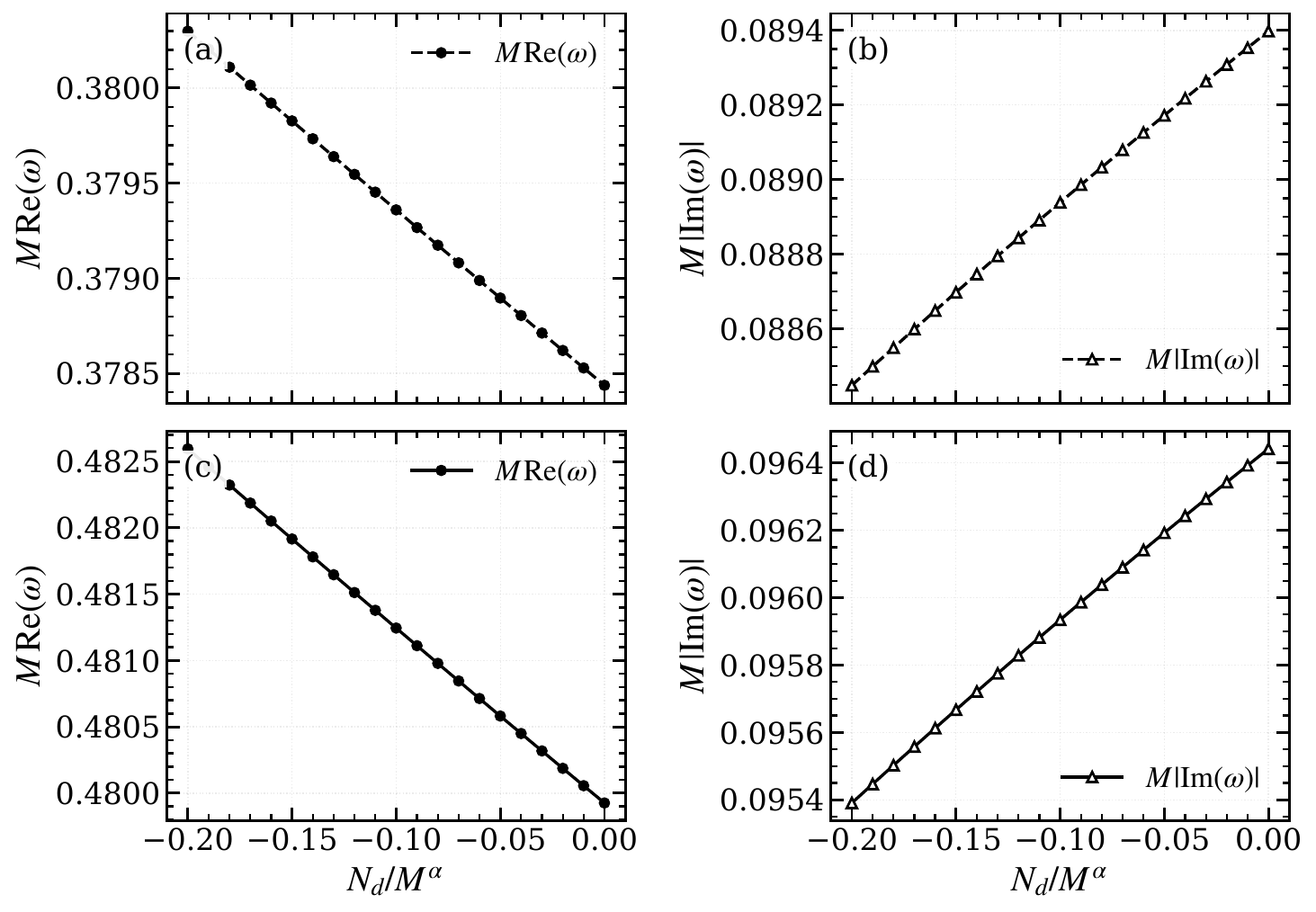}}
    \end{center}
    \caption{
    The dependence of the dimensionless fundamental
    quasinormal frequencies \(M\omega\)
    on the dimensionless surrounding-matter parameter \(N_d/M^\alpha\)
    in the large-\(\gamma\) region.
    The parameters are fixed as \(e/M=0.4\) and \(\gamma=0.5\).
    For the \(\ell=2\) case,
    the upper two panels correspond to the GR-led branch,
    while the lower two panels correspond to the EM-led branch.
    }
    \label{fig:qnm-Nd-large}
\end{figure*}

For \(\gamma=0.5\),
we consider negative values of \(N_d/M^\alpha\).
As shown in
Fig.~\ref{fig:qnm-Nd-large},
\(M\operatorname{Re}\omega\)
decreases as \(N_d/M^\alpha\) approaches zero,
while
\(M|\operatorname{Im}\omega|\)
increases.
Equivalently,
as \(|N_d|/M^\alpha\) increases,
the oscillation frequency increases
and the damping rate decreases.
This behavior occurs for the \(\ell=1\) mode
and for both \(\ell=2\) branches.
Taken together,
the two parameter regions show that
the influence of \(N_d/M^\alpha\) depends on the
Rastall-coupling region.
In the small-\(\gamma\) region,
increasing positive \(N_d/M^\alpha\)
lowers both the oscillation frequency
and the damping rate.
In the large-\(\gamma\) region,
increasing the magnitude of the negative
\(N_d/M^\alpha\) raises the oscillation frequency
while reducing the damping rate.

\section{Conclusion} \label{conclusion}

We have established a compatible formulation of axial
gravitoelectromagnetic perturbations for charged black holes
surrounded by dust-like Kiselev-type anisotropic matter in Rastall
gravity. The linearized matter equation yields a compatibility
condition linking the source and metric perturbations. With the
closure $\delta s_\mu=0$, the induced covariant source terms satisfy
this condition identically, and the gravitational and electromagnetic
perturbations reduce to two coupled Schr\"odinger-type master equations.

We determined the algebraic decoupling properties of this system.
The neutral limit separates the two sectors, while charged
configurations admit a constant decoupling in the RN and
general-relativistic dust-like limits. In a generic charged Rastall
background with nonzero surrounding matter, additional radial
structure obstructs a constant transformation to independent master
equations. The physical axial dipole follows separately from the
radial constraint: its electromagnetic component evolves autonomously
and induces a constrained metric response.

We proved axial mode stability for the adopted matter response on
the nonextremal backgrounds specified in
Sec.~\ref{subsec:axial_mode_stability}.
For $\ell\geq2$, a constant symmetrization and an explicit matrix
$S$-deformation yield a positive-definite deformed potential throughout
the static exterior. The resulting integral identity excludes
exponentially growing modes without requiring algebraic decoupling.
The positive effective potential of the physical electromagnetic
dipole establishes the corresponding $\ell=1$ result.
The proof covers all physical axial multipoles and overtone numbers,
independently of the numerical parameter sampling.

Using a Chebyshev pseudospectral method, validated against the RN
spectrum and convergence with spectral order, we determined the
fundamental frequencies and their dependence on the background
parameters. In the ranges examined, increasing the Rastall coupling
lowers both the oscillation frequency and the damping rate in the
branch connected to general relativity. In the large-\(\gamma\) region,
the oscillation frequency increases, while the damping rate varies
nonmonotonically.
Over the scanned ranges, the oscillation frequency increases
monotonically with \(e/M\) at fixed \(\gamma\) and \(N_d/M^\alpha\),
while the damping rate varies nonmonotonically.
At fixed \(\gamma\) and \(e/M\), increasing \(|N_d|/M^\alpha\)
reduces the damping rate in both regions.
All computed fundamental frequencies have
negative imaginary parts, consistently with the analytic
mode-stability result.

This work establishes a constraint-compatible axial perturbation
framework, proves mode stability under the adopted closure,
and determines the fundamental-mode frequencies.
\begin{acknowledgments}
This work was supported by the National Natural Science Foundation of China (Grants
No. 12205129 and No. 12247101), the Fundamental Research Funds for the Central Universities (Grants No. lzujbky-2025-it05 and lzujbky-2025-jdzx07), the Natural Science 
Foundation of Gansu Province (No. 22JR5RA389, No. 25JRRA799), the ‘111 Center’ under Grant
No. B20063, and the Key Project of the Department of Education of Hunan Province (No.
25A0084). Wen-Di Guo was supported by “Talent Scientific Fund of Lanzhou University”.
\end{acknowledgments}


\appendix
\section{Component equations and constraint propagation}
\label{app:axial_constraints}

We give the axial component equations for the source amplitudes
$\tau_t$ and $\tau_r$ defined in
Eq.~\eqref{eq:axial_source_amplitudes} and derive the identity that
ensures consistency of the master-equation reduction. We use
$\omega\neq0$ and the matter compatibility condition
Eq.~\eqref{eq:axial_matter_constraint}.

Define
\begin{equation}
P(r)=\kappa p_t(r),
\qquad
\mathcal Z=h_0'-\frac{2h_0}{r}+\mathrm i\omega h_1
-\frac{4e}{r^2}f_M,
\qquad
\mathcal C_g=\mathrm i\omega h_0+f(fh_1)'.
\label{eq:axial_auxiliary_variables}
\end{equation}
Let $\mathcal E_t$ and $\mathcal E_r$ be the radial coefficients of
the $t\phi$ and $r\phi$ residuals of
Eq.~\eqref{eq:linearized_rastall_equation}, with all terms moved to
the left-hand side and the factor
$\mathcal S_\ell\mathrm e^{-\mathrm i\omega t}$ removed.
Using the background equations, these coefficients are
\begin{align}
\mathcal E_r&=\frac{\mathrm i\omega}{2f}\mathcal Z
+\frac{K_\ell}{2r^2}h_1+P h_1-\kappa\tau_r,
\label{eq:axial_radial_equation}\\
\mathcal E_t&=-\frac{f}{2r^2}(r^2\mathcal Z)'
+\frac{K_\ell}{2r^2}h_0+P h_0-\kappa\tau_t.
\label{eq:axial_temporal_equation}
\end{align}
The full equations require $\mathcal E_r=\mathcal E_t=0$.
The angular residual and the Maxwell equation take the forms
\begin{align}
\mathcal E_{\theta\phi}
&=\frac{\partial_\theta\mathcal S_\ell
-2\cot\theta\,\mathcal S_\ell}{2f}
\mathcal C_g\,\mathrm e^{-\mathrm i\omega t},
\label{eq:axial_angular_equation}\\
0&=\frac{d^2 f_M}{dr_*^2}
+\left[\omega^2-f\left(\frac{L_\ell}{r^2}
+\frac{4e^2}{r^4}\right)\right]f_M
-\frac{ef}{r^2}\mathcal Z.
\label{eq:axial_maxwell_equation}
\end{align}
For $\ell\geq2$, the angular equation requires $\mathcal C_g=0$,
or $h_0=\mathrm i f(fh_1)'/\omega$.

Differentiating the radial residual and using
Eq.~\eqref{eq:axial_matter_constraint} yields the identity
\begin{equation}
\mathcal E_t=
\frac{\mathrm i f}{\omega r^2}(r^2 f\mathcal E_r)'
+\frac{K_\ell}{2\mathrm i\omega r^2}\mathcal C_g
+\frac{\mathrm i\kappa f}{\omega}\mathcal C_m.
\label{eq:axial_constraint_identity}
\end{equation}

For the matter response $\tau_t=p_t h_0$ and $\tau_r=p_t h_1$,
the matter constraint vanishes identically. For $\ell\geq2$, the
radial and angular equations then imply the temporal equation
through Eq.~\eqref{eq:axial_constraint_identity}.
For $\ell=1$, the angular residual vanishes identically and the
radial equation gives $\mathcal Z=0$. The physical dipole equation
follows from the Maxwell equation, as derived in
Sec.~\ref{subsec:physical_axial_dipole}.

\bibliographystyle{apsrev4-2}
\bibliography{references}

\end{document}